\documentclass[11pt]{article}

\usepackage[final]{acl}

\usepackage{times}
\usepackage{latexsym}

\usepackage[T1]{fontenc}

\usepackage[utf8]{inputenc}

\usepackage{microtype}

\usepackage{inconsolata}

\usepackage{breakcites}

\usepackage{graphicx}
\usepackage{float}
\usepackage{enumitem}
\usepackage{booktabs}

\usepackage{pgfplots}
\usepackage{caption}
\usepackage{framed}
\usepackage{ragged2e}
\usepackage{makecell}

\usepackage{cite}
\usepackage{amsmath,amssymb,amsfonts}
\usepackage{textcomp}
\usepackage{xcolor}
\usepackage{xspace}
\usepackage{balance}
\usepackage{multirow}
\usepackage{hyperref}
\usepackage{pifont}
\usepackage{listings}
\usepackage{adjustbox}
\usepackage{colortbl}
\usepackage{amsmath}
\usepackage{amssymb}
\usepackage{tikz}

\makeatletter
\def\endthebibliography{%
  \def\@noitemerr{\@latex@warning{Empty `thebibliography' environment}}%
  \endlist
}
\makeatother

\newenvironment{result}{\begin{framed}\centering\it}{\end{framed}}

\newcommand{\recheck}[1]{\textcolor{black}{#1}}
\newcommand{\revise}[1]{\textcolor{black}{#1}}

\newcommand{\camcheck}[1]{\textcolor{black}{#1}}

\newcommand{\approach}{\textsc{MuCoCo}\xspace}
\newcommand{\turb}{\textsc{Turbulence}\xspace}
\newcommand{\cmark}{{\color{green}\ding{51}}}  
\newcommand{\xmark}{{\color{red}\ding{55}}} 

\newcommand{\ccmark}{{\color{black}\textcircled{\checkmark}}}
\newcommand{\cxmark}{{\color{black}$\otimes$}}

\newcommand*{\TakeFourierOrnament}[1]{{%
\fontencoding{U}\fontfamily{futs}\selectfont\char#1}}
\newcommand*{\inv}{{\footnotesize \TakeFourierOrnament{66}}}

\newcommand{\fullcircle}{\begin{tikzpicture}[baseline=-0.5ex]\draw[fill=black] (0,0) circle (0.5ex);\end{tikzpicture}}
\newcommand{\halfcircle}{\begin{tikzpicture}[baseline=-0.5ex]\draw (0,0) circle (0.5ex); \fill (0,0) -- (90:0.5ex) arc (90:270:0.5ex) -- cycle;\end{tikzpicture}}
\newcommand{\emptycircle}{\begin{tikzpicture}[baseline=-0.5ex]\draw (0,0) circle (0.5ex);\end{tikzpicture}}

\definecolor{codegreen}{rgb}{0,0.6,0}

\def\BibTeX{{\rm B\kern-.05em{\sc i\kern-.025em b}\kern-.08em
    T\kern-.1667em\lower.7ex\hbox{E}\kern-.125emX}}
\begin{document}

\title{\approach: 
Automated Consistency Testing 
of Code LLMs
}



\author{Chua Jin Chou$^1$ \quad Khant That Lwin$^2$ \quad Ezekiel Soremekun$^1$ \\
  $^1$Singapore University of Technology and Design (SUTD) \\
  $^2$Royal Holloway University of London (RHUL) \\
  \texttt{\{jinchou\_chua, ezekiel\_soremekun\}@sutd.edu.sg} \quad \texttt{Khant.Lwin.2023@live.rhul.ac.uk} \\
}
\setlength\titlebox{7cm}

\maketitle
\begin{abstract}
Code LLMs often portray inconsistent program behaviors. 
Developers typically employ benchmarks to assess Code LLMs, but most benchmarks are hand-crafted,  static and do not target \textit{consistency} property. In this work, we pose the scientific question:\textit{ how can we automatically discover inconsistent program behaviors in Code LLMs?} To address this challenge, we propose an automated consistency testing method, called \approach, which employs 
\textit{semantic-preserving 
mutation analysis} to expose inconsistent behaviors in code LLMs. Given a coding query, 
\approach 
automatically transforms a program into semantically equivalent programs 
(aka mutants) 
and detects inconsistencies between the mutants and the original program (e.g., different output or test failure). 
We evaluate \approach using 
\recheck{four (4)} coding tasks 
and \recheck{seven (7)} LLMs. 
Results show that 
\approach is effective in exposing inconsistency and outperforms the closest baseline (\turb).  About \recheck{one in seven (15\%)} inputs generated by \approach exposed inconsistencies. 
Our work 
motivates the need to test Code LLMs for consistency property. 
\end{abstract}

\section{Introduction}
\label{sec:intro}

Software practitioners are increasingly employing Code Large Language Models (Code LLMs) for programming and software engineering (SE) tasks.  
Similarly, researchers are increasingly deploying LLMs to address SE tasks, such as code generation and execution prediction~\citep{fakhoury2024llm}, 
with remarkable performance
~\citep{bouzenia2025repairagent, jin2023inferfix, 10.1145/3712003}. 
Given the increasingly high reliance on LLMs for coding tasks,  it is important to assess the performance of LLMs and ensure that they are reliable and consistent.  
Particularly,  lack of consistency may lead to fatal consequences in critical use cases. 

Practitioners currently employ manually curated coding benchmarks to assess Code LLMs. 
Several benchmarks have been proposed to measure the effectiveness of Code LLMs,  including HumanEval~\citep{chen2021evaluating},  MBPP~\citep{austin2021program},  CodeMMLU~\citep{nguyen2025codemmlu},  CruxEval~\citep{pmlr-v235-gu24c} and BigCodeBench~\citep{zhuo2025bigcodebench}.   Likewise,  \recheck{many leaderboards measure/rank LLMs performance on varying coding tasks}~\citep{evalplus_leaderboard, llm_stats_humaneval, bigcode_bench,  zhang2025swebenchgoeslive, evalooop}. 

However,  existing benchmarks are manually curated and focus on measuring the \textit{correctness} of LLMs \citep{guan2026benchmarkusefuldynamicbenchmarking, chen2021evaluating, zhuo2025bigcodebench, pmlr-v235-gu24c, nguyen2025codemmlu, austin2021program}. 
Manual benchmark curation is challenging and these benchmarks do not apply to consistency property.  
Besides, benchmarks quickly leak to newer generation of LLMs over time~\citep{guan2026benchmarkusefuldynamicbenchmarking}. 

Beyond correctness,  it is important to assess non-functional properties of Code LLMs like \textit{consistency},  since LLMs often produce \textit{inconsistent} results across multiple runs, varying prompts or even similar tasks \citep{rajan-etal-2024-knowledge}.  \textit{Inconsistent} LLM behaviors may have severe consequences during 
programming.  
Researchers have observed that Code LLMs may generate incorrect programs 
for similar 
code queries,  syntax or logic \citep{guan2026benchmarkusefuldynamicbenchmarking}. 
\autoref{tab:mucoco-random-only} highlights an example of 
inconsistent behaviors in state-of-the-art Code LLMs.

\definecolor{codegreen}{rgb}{0,0.6,0}
\definecolor{codegray}{rgb}{0.5,0.5,0.5}
\definecolor{codepurple}{rgb}{0.58,0,0.82}
\definecolor{backcolour}{rgb}{0.95,0.95,0.92}

\lstdefinestyle{mystyle}{
    backgroundcolor=\color{backcolour},   
    commentstyle=\color{codegreen},
    keywordstyle=\color{magenta},
    numberstyle=\tiny\color{codegray},
    stringstyle=\color{codepurple},
    basicstyle=\ttfamily\footnotesize,
    breakatwhitespace=false,         
    breaklines=true,                 
    captionpos=b,                    
    keepspaces=true,                 
    numbers=left,                    
    numbersep=6pt,                  
    xleftmargin=2em,                
    showspaces=false,                
    showstringspaces=false,
    showtabs=false,                  
    tabsize=2
}

\lstset{style=mystyle}

\begin{table*}[tb!]
\centering
\caption{
\centering
\textsc{Sample consistency test generated by MuCoCo and its
LLM Outputs.  
(\cmark = consistency, \xmark = inconsistency,
OG = LLM Answer for Original Question, MU = LLM Answer for Mutated Question).
Appendix (\autoref{tab:mucoco-examples}) provides 
more examples for each mutation type.}}
\label{tab:mucoco-random-only}
\begin{adjustbox}{max width=\textwidth}
\renewcommand{\arraystretch}{1.1}
\setlength{\tabcolsep}{3pt}
\begin{tabular}{c|l|l|c|c|c|c|c|c|c|c}
\toprule
\textbf{Mutation Type} &
\multicolumn{1}{c|}{\textbf{Original Question}} &
\multicolumn{1}{c|}{\textbf{Mutated Question}} &
\begin{tabular}[c]{@{}l@{}}\textbf{Dataset}\\\textbf{(Canon. Ans.)}\end{tabular} &
\textbf{\begin{tabular}[c]{@{}c@{}} GPT-4o \\ (OG) \\ (MU) \end{tabular}} &
\textbf{\begin{tabular}[c]{@{}c@{}} GPT-5 \\ (OG) \\ (MU) \end{tabular}} &
\textbf{\begin{tabular}[c]{@{}c@{}} Qwen \\ (OG) \\ (MU) \end{tabular}} &
\textbf{\begin{tabular}[c]{@{}c@{}} Llama \\ (OG) \\ (MU) \end{tabular}} &
\textbf{\begin{tabular}[c]{@{}c@{}} Gemma \\ (OG) \\ (MU) \end{tabular}} &
\textbf{\begin{tabular}[c]{@{}c@{}} DeepSeek \\ (OG) \\ (MU) \end{tabular}} &
\textbf{\begin{tabular}[c]{@{}c@{}} Codestral \\ (OG) \\ (MU) \end{tabular}} \\
\midrule

\textbf{Random} &
\begin{minipage}{0.35\linewidth}
\vspace{-6pt}
\begin{lstlisting}[language=Python]
def f(text, n):
  if n < 0 or len(text) <= n:
    return text
  result = text[0:n]
  i = len(result) - 1
  while i >= 0:
    if result[i] != text[i]:
      break
    i -= 1
  return text[0:i + 1]
\end{lstlisting}
\vspace{-4pt}
\end{minipage}
&
\begin{minipage}{0.45\linewidth}
\vspace{-6pt}
\begin{lstlisting}[language=Python]
def dOdtZZSuisMB(TYesQRBm, mXDbZtJMZFV):
  if mXDbZtJMZFV < 0 or len(TYesQRBm) <= mXDbZtJMZFV:
    return TYesQRBm
  result = TYesQRBm[0:mXDbZtJMZFV]
  i = len(result) - 1
  while i >= 0:
    if result[i] != TYesQRBm[i]:
      break
    i -= 1
  return TYesQRBm[0:i + 1]
\end{lstlisting}
\vspace{-4pt}
\end{minipage}
&
\begin{tabular}[c]{@{}c@{}}CruxEval\_789 \\ ('br')\end{tabular}
&
\begin{tabular}[c]{@{}c@{}} \xmark \\ ('br') \\ (['bR']) \end{tabular}
&
\cmark
&
\begin{tabular}[c]{@{}c@{}} \xmark \\ ('br') \\ (['bR']) \end{tabular}
&
\cmark
&
\begin{tabular}[c]{@{}c@{}} \xmark \\ ('br') \\ (['bR']) \end{tabular}
&
\begin{tabular}[c]{@{}c@{}} \xmark \\ ('br') \\ (['bR']) \end{tabular}
&
\begin{tabular}[c]{@{}c@{}} \xmark \\ ('br') \\ (['bR']) \end{tabular}
\\

\bottomrule
\end{tabular}
\end{adjustbox}
\end{table*}

To this end,  this work poses the following scientific question: \textit{How can we automatically assess the consistency property of Code LLMs?} We propose an automated technique for assessing the consistency of Code LLMs called \approach\footnote{\approach = \textbf{Mu}tation-based \textbf{Co}de \textbf{Co}nsistency Testing.}. The main idea of our approach is to employ \textit{semantic-preserving mutation analysis} 
to discover inconsistency in Code LLMs.  

\begin{figure}
  \includegraphics[width=\columnwidth]{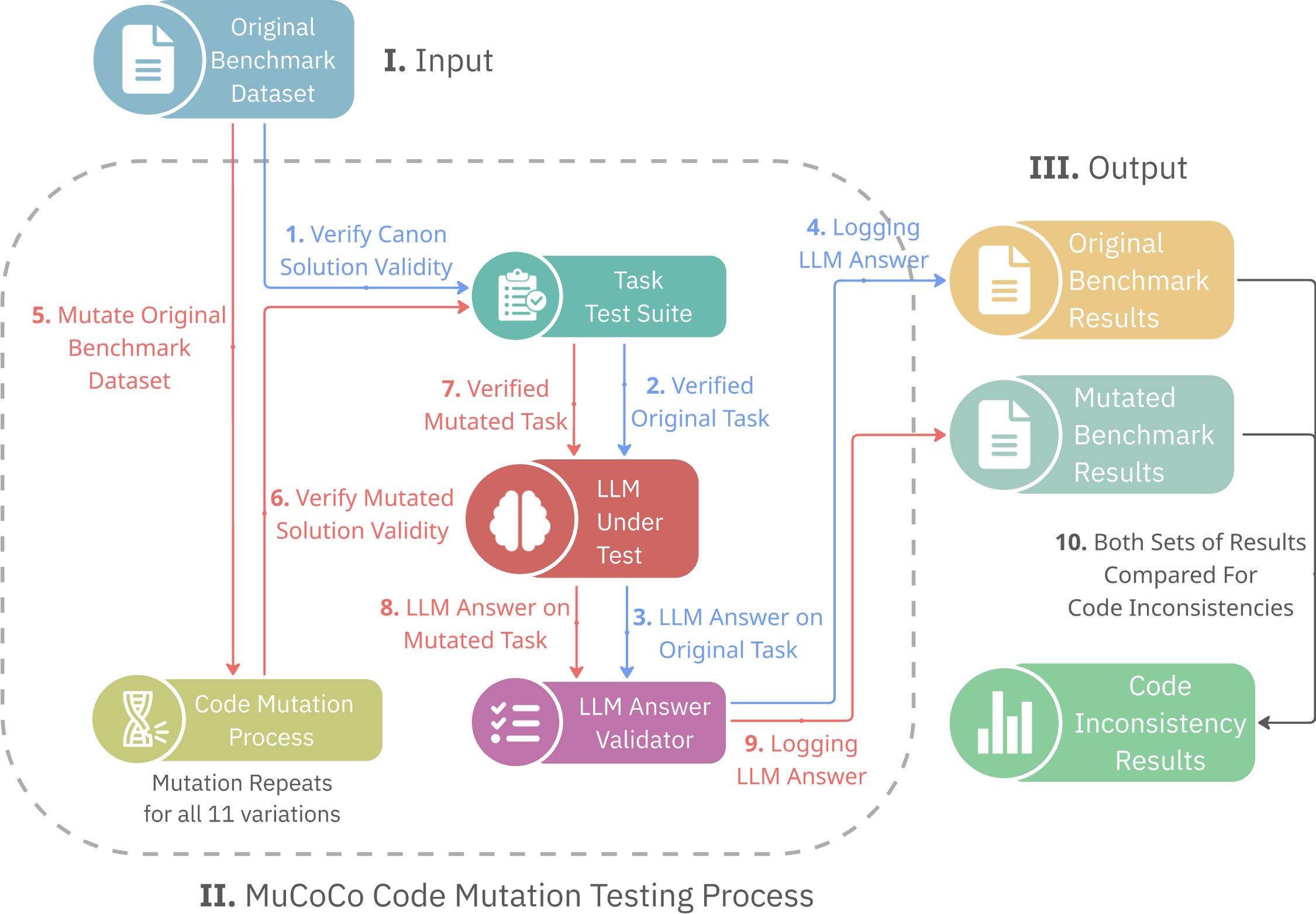}
  \caption{\approach Workflow}
  \label{fig:workflow}
\end{figure}

\autoref{fig:workflow} illustrates the workflow of \approach: Given an original dataset (e.g., HumanEval \citep{chen2021evaluating}),  \approach automatically mutates its code queries into semantically equivalent queries. 
\recheck{Then, its metamorphic oracle detects consistency error (aka inconsistency) by examining the test correctness of the LLM outputs for the original query versus the mutated query.  It discovers inconsistency when the LLM outputs of both queries differ, e.g., when the output of the mutated query fails a test but the output of the original passes, or vice versa}.  
For instance, \autoref{tab:mucoco-random-only}
shows that \approach's random renaming of CruxEval-789 \citep{pmlr-v235-gu24c} triggered consistency errors (\xmark) in 
 five LLMs (GPT-4o, 
Qwen,  
DeepSeek,  
and CodeStral).  

In summary, our contributions are as follows: 
\begin{enumerate} [leftmargin=*]
\item We propose an automated method (\approach) that employs mutation analysis and metamophic testing to discover \textit{in}consistency in Code LLMs. \approach provides \recheck{11} semantic-preserving mutations (\textit{see} \autoref{sec:methodology} and \autoref{tab:mucoco-examples}). 

\item We evaluate \approach (\autoref{sec:experiment-setup}) using \recheck{seven} state-of-the-art LLMs and \recheck{four}  benchmarks.  
We found that \recheck{14.82}\% (\recheck{22K/148K}) of inputs generated by \approach reveals \textit{in}consistency in Code LLMs (\textbf{RQ1},  \autoref{sec:results}).  

\item  We compare \approach to a hand-crafted benchmark -- 
\turb~\citep{10989005}.  Results demonstrate that \approach is up to \recheck{six (6)} times more effective than \turb in consistency testing,  and 
discovers new classes of consistency errors (\textbf{RQ2}, \autoref{sec:results}). 
\end{enumerate}

\section{Overview}
\label{sec:overview}

%
\noindent
\recheck{\textbf{Code Consistency Property:} 
\autoref{tab:inconsistency-attribution} highlights the types of \textit{in}consistency detected by \approach.  Given a \textit{consistency test} (a pair of semantically-equivalent code queries),  we say that an LLM is \textit{inconsistent} (\xmark) if one of these conditions is met:}

\begin{itemize} [leftmargin=*]

\item the LLM output is \textit{correct} (\ccmark) for a coding query,  but \textit{incorrect} (\cxmark) for an equivalent query (aka \textit{correctness-based inconsistency}).

\item  the LLM output  is \textit{correct}  (\ccmark) or \textit{incorrect} (\cxmark) for a query,   but \textit{invalid} (\inv)
for an equivalent query (aka \textit{invalidity-based inconsistency}).

\item LLM output for \textit{both} queries are \textit{incorrect} (\cxmark), albeit for different reasons, e.g.,  failed different tests (aka \textit{incorrectness-based inconsistency}). 

\end{itemize}

\recheck{
Analogously,  we say an LLM is \textit{consistent} (\cmark) for a pair of semantically-equivalent code queries,  if both outputs are correct (\ccmark),  incorrect (\cxmark) or invalid (\inv) for the same reasons, i.e.,  there is no test case that distinguishes both outputs. }
\recheck{
Consistency property is \textit{different} from  \textit{correctness} property: Unlike traditional model performance properties (e.g., accuracy),  \textit{(in)consistency} does not \textit{require}  LLM outputs to be correct.  
In contrast to \textit{robustness} property~\citep{10989005},  \textit{(in)consistency} property requires that code queries are \textit{semantically-equivalent} but produce different test outcomes. 
}


We determine \textit{correctness}  by (a) comparing the LLM output to the expected output, 
and (b) ensuring the LLM-generated outputs pass the test suite. 
Concretely,  \textit{correctness} (\ccmark) means that the LLM output matches the expected output and passes all test cases.  \textit{Incorrectness} (\cxmark) means the LLM output does not match the expected output or fails at least one test case. 
\textit{Invalidity} (\inv) means the LLM output violates the expected output specification, e.g., non-executable program or an empty output.  
An in-depth explanation of our correctness and inconsistency oracles is provided in Section \ref{sub:test-oracle}.


\smallskip
\noindent
\textbf{Key Insight:}
To automatically discover consistency errors,  we propose an automated testing technique called  \approach.  The main idea of our technique is to employ \textit{mutational analysis} and \textit{metamorphic testing} to detect consistency errors in Code LLMs.  The key insight of \approach is that 
\begin{center}
\textit{a pair of code queries with the same semantics (meaning) should produce LLM outputs that exhibit similar behaviors}. 
\end{center}

In particular, we posit that given a pair of semantically equivalent code queries,  the LLM should produce a pair of outputs that exhibit similar behaviors,  i.e.,  same outputs and test correctness.  We hypothesize that if a pair of equivalent code queries produce differing LLM outputs or behaviors, then the LLM is inconsistent in handling the query.  

Leveraging this metamorphic property,  we design \approach to generate semantic-preserving code queries. 
Given an existing LLM input (e.g., prompt or code from HumanEval),  \approach mutates it into a semantic-preserving equivalent query, feeds them into the LLM and examines whether the behavior of the LLM is similar for both queries.  
\approach employs lexical,  syntactic and logical mutations that preserve the program semantics, e.g.,  variable renaming,  for-to-while loop transformation and constant folding \citep{guan2026benchmarkusefuldynamicbenchmarking} \citep{orvalho2026largelanguagemodelsrobust},  respectively.  

\begin{table}[h]
\centering
\tiny
\caption{\centering
  \textsc{Consistency Errors based on LLM Outcomes
    (OG = Original,  Mut = Mutated).} 
}
\renewcommand{\arraystretch}{1}
\setlength{\tabcolsep}{6pt}
\resizebox{\columnwidth}{!}{\tiny
\begin{tabular}{llc}
\toprule
\textbf{OG Task} & \textbf{Mut Task} & \textbf{(\textit{In})Consistency} \\
\midrule
Correct   & Incorrect & \multirow{2}{*}{\begin{tabular}[c]{@{}c@{}}Correctness-based\\ Inconsistency (\xmark)\end{tabular}}  \\
Incorrect & Correct   &                                   \\
\midrule
Incorrect & Incorrect & \multirow{2}{*}{\begin{tabular}[c]{@{}c@{}}Incorrectness-based\\ Inconsistency (\xmark)\end{tabular}} \\
\multicolumn{2}{c}{\textit{(different test outcomes)}} &   \\
\midrule

Correct   & Invalid   & \multirow{4}{*}{\begin{tabular}[c]{@{}l@{}}Invalidity-based\\ Inconsistency (\xmark)\end{tabular}} \\
Incorrect & Invalid   &                                 \\
Invalid   & Incorrect &                                 \\
Invalid   & Correct   &                                 \\
\midrule
Correct   & Correct   & \textit{Consistent} (\cmark)\\
Invalid   & Invalid   & \textit{Consistent} (\cmark) \\
\midrule 
Incorrect & Incorrect &  \multirow{2}{*}{\textit{Consistent} (\cmark)}  \\
\multicolumn{2}{c}{\textit{(similar test outcomes)}}  &  \\

\bottomrule
\end{tabular}
}
\label{tab:inconsistency-attribution}
\end{table}

\begin{table*}[tb!]
\caption{\centering \textsc{\camcheck{
Detailed Comparison of 
\approach against SOTA frameworks.  \protect\fullcircle\ implies "fully", \protect\halfcircle\ implies "partially" and \protect\emptycircle\ implies "does not" support the specified feature.}}}\centering
\resizebox{\textwidth}{!}{
\begin{tabular}{|c|c|c|c|c|c|c|c|c|c|c|c|c|c|c|}
\hline
\multirow{2}{*}{Framework}
 & \multirow{3}{*}{\begin{tabular}{c}Mutation\\Testing\end{tabular}}
 & \multirow{3}{*}{\begin{tabular}{c}Automatic\\Testing\\Framework\end{tabular}}
 & \multicolumn{2}{c|}{LLM Benchmarking}
 & \multicolumn{2}{c|}{Test Oracles}
 & \multicolumn{4}{c|}{Coding Tasks}
 & \multicolumn{3}{c|}{Testing Goals}
 & \multirow{3}{*}{\begin{tabular}{c}Natural\\Language\end{tabular}}
\\
\cline{4-14}
 &
 &
 & \begin{tabular}{c}Automated\\Generation\end{tabular} & \begin{tabular}{c}Manual\\Generation\end{tabular}
 & Automated & Metamorphic
 & \begin{tabular}{c}Code\\ Gen.\end{tabular} & \begin{tabular}{c}Input\\ Pred.\end{tabular} & \begin{tabular}{c}Output\\ Pred.\end{tabular} & MCQ 
 & Consistency & Robustness & Accuracy
 &
\\
\hline

\approach (our approach)
 & \fullcircle 
 & \fullcircle
 & \emptycircle & \emptycircle
 & \fullcircle & \fullcircle 
 & \fullcircle & \fullcircle & \fullcircle & \fullcircle
 & \fullcircle & \halfcircle & \halfcircle
 & \emptycircle
\\
\hline

Turbulence
\citep{10989005}
 & \emptycircle 
 & \emptycircle
 & \halfcircle & \fullcircle
 & \fullcircle & \emptycircle 
 & \fullcircle & \emptycircle & \emptycircle & \emptycircle
 & \halfcircle & \fullcircle & \halfcircle
 & \emptycircle
\\
\hline

KonTest
\citep{rajan-etal-2024-knowledge}
 & \fullcircle 
 & \fullcircle
 & \emptycircle & \emptycircle
 & \fullcircle & \fullcircle 
 & \emptycircle & \emptycircle & \emptycircle & \emptycircle
 & \fullcircle & \halfcircle & \halfcircle
 & \fullcircle
\\
\hline

\citet{guan2026benchmarkusefuldynamicbenchmarking}
 & \fullcircle & \emptycircle
 & \fullcircle & \emptycircle 
 & \fullcircle & \emptycircle
 & \halfcircle & \emptycircle & \fullcircle & \emptycircle
 & \halfcircle & \fullcircle & \halfcircle
 & \emptycircle
\\
\hline

DyCodeEval
\citep{3780338.3780675}
 & \emptycircle & \emptycircle
 & \fullcircle & \emptycircle 
 & \fullcircle & \emptycircle
 & \fullcircle & \emptycircle & \emptycircle & \emptycircle
 & \halfcircle & \fullcircle & \halfcircle
 & \fullcircle
\\
\hline

DynaCode
\citep{hu-etal-2025-dynacode}
 & \emptycircle & \emptycircle
 & \fullcircle & \emptycircle 
 & \fullcircle & \emptycircle
 & \fullcircle & \emptycircle & \emptycircle & \emptycircle
 & \halfcircle & \fullcircle & \halfcircle
 & \emptycircle
\\
\hline

PPM
\citep{10.1145/3643780}
 & \emptycircle & \emptycircle
 & \fullcircle & \emptycircle 
 & \fullcircle & \emptycircle
 & \fullcircle & \emptycircle & \emptycircle & \emptycircle
 & \halfcircle & \fullcircle & \halfcircle
 & \emptycircle
\\
\hline

EquiBench
\citep{wei-etal-2025-equibench}
 & \fullcircle & \emptycircle
 & \fullcircle & \emptycircle 
 & \fullcircle & \emptycircle
 & \emptycircle & \emptycircle & \halfcircle & \emptycircle
 & \emptycircle & \emptycircle & \fullcircle
 & \emptycircle
\\
\hline

CodeCrash
\citep{NEURIPS2025_aec5e284}
 & \fullcircle & \fullcircle
 & \fullcircle & \emptycircle 
 & \fullcircle & \emptycircle
 & \emptycircle & \fullcircle & \fullcircle & \emptycircle
 & \halfcircle & \fullcircle & \halfcircle
 & \fullcircle
\\
\hline

CCTest
\citep{10172845}
 & \fullcircle & \fullcircle
 & \emptycircle & \emptycircle 
 & \fullcircle & \fullcircle
 & \fullcircle & \emptycircle & \emptycircle & \emptycircle
 & \fullcircle & \halfcircle & \halfcircle
 & \emptycircle
\\
\hline

\citet{orvalho2026largelanguagemodelsrobust}
 & \fullcircle & \emptycircle
 & \emptycircle & \emptycircle 
 & \halfcircle & \emptycircle
 & \emptycircle & \emptycircle & \fullcircle & \emptycircle
 & \halfcircle & \fullcircle & \halfcircle
 & \emptycircle
\\
\hline

\end{tabular}
}
\label{tab:sota-mucoco-comparison}
\end{table*}

\smallskip
\noindent
\textbf{Novelty w.r.t. the State-of-the-art: }
To the best of our knowledge,  \approach is the \textit{first} automated method for assessing the \textit{consistency} of Code LLMs.  
\approach is unique with its combination of semantic-preserving mutations and metamorphic testing, and its focus on programming tasks. 
Specifically, \approach is the first approach that targets consistency property in Code LLMs. 
It 
does not require manual effort/intervention and 
applies to arbitrary code benchmarks. 

Existing benchmarks are manually curated~\citep{zhuo2025bigcodebench, pmlr-v235-gu24c, nguyen2025codemmlu} and do not apply to \textit{consistency} testing.  They often focus on the \textit{correctness} of LLMs.  
Besides,  static LLM benchmarks quickly leaked to newer models~\citep{guan2026benchmarkusefuldynamicbenchmarking}. 
Existing dynamic benchmarks~\citep{guan2026benchmarkusefuldynamicbenchmarking}  
target the \textit{correctness} property,  while \approach focuses on code consistency.
\turb~\citep{10989005}, provides program templates that can be instantiated to discover robustness errors in LLMs. 
\turb is the closest related work to our technique,  but it is manually crafted and focuses on model robustness instead of consistency testing like \approach.
In this work, we compare \approach to \turb and illustrate that \approach outperforms and complements \turb 
(\textit{see} \textbf{RQ2}  \autoref{sec:results}). 
Similar to \approach, KonTest~\citep{rajan-etal-2024-knowledge} employs mutation analysis and metamorphic properties to automatically assess consistency, albeit in text-based LLMs, it   
does not apply to Code LLMs. 

\camcheck{
%
The state-of-the-art (SOTA) automatic mutation testing frameworks for Code LLMs 
includes CodeCrash \citep{NEURIPS2025_aec5e284} and CCTest \citep{10172845}.
CodeCrash \citep{NEURIPS2025_aec5e284} evaluates the reasoning reliability of LLMs through perturbations and misleading natural language prompts.
Conversely, \approach focuses on code-related perturbations,  it neither perturbs the natural language channel nor examine 
model reasoning for inconsistencies. 
CCTest \citep{10172845} is a metamorphic testing framework which mutates incomplete program inputs using a suite of mutation strategies to test and repair code completion systems. 
While CCTest \citep{10172845} evaluates inconsistencies in model outputs between mutated programs, their approach involves measuring the 
code similarity  between two programs  using the BLEU score and 
Levenshtein distance. 
This approach differs from \approach, which measures inconsistencies in code generation tasks by first checking the correctness of both the original and mutated outputs against the benchmark's test suite before comparing them to identify inconsistent behaviour.
Furthermore,  \approach supports the largest variety of coding tasks such as input prediction, output prediction and MCQ tasks. These tasks are not supported by SOTA methods such as CCTest \citep{10172845}.
\autoref{tab:sota-mucoco-comparison} provides a comprehensive comparison of \approach against SOTA benchmarks and frameworks.  Additionally, \autoref{tab:sota-mucoco-mutation-comparison} compares the mutation operators supported by \approach versus SOTA approaches.}

\section{Methodology}
\label{sec:methodology}




\autoref{fig:workflow} highlights \approach's workflow.

\smallskip
\noindent
\textbf{1) Original Benchmark Testing:}
\approach's consistency testing starts with retrieving the original benchmark dataset (e.g., HumanEval). 
Next,  for each task in the benchmark,  \approach executes its canonical solution using the benchmark's test suite 
to ensure dataset correctness. The LLM under test is fed the task,  and its output is collected 
and stored.  Then,  \approach executes the benchmark's test suite on the LLM output and stores the results. 
\autoref{fig:workflow} highlights these steps in {\color{blue} blue} lines.

\smallskip
\noindent
\textbf{2) Mutated Benchmark Testing:}
\approach mutates each task in the original benchmark (e.g., HumanEval/1) for each semantic-preserving 
mutation operator. 
First,  \approach checks that the mutation operator is possible for the task at hand, e.g.,  it checks that there is a \texttt{for} loop in the program during a \texttt{for-to-while} mutation.  
It then mutates the task at hand using the desired mutation operator and the mutated task is tested against its corresponding test suite 
to ensure that correctness is preserved post-mutation.
Next,  the LLM is fed the mutated task and 
its results are collected and stored.
These steps are outlined in \autoref{fig:workflow} using {\color{red} red} lines.

\begin{table*}[tb!] 
\caption{\centering Details of Tested LLM Models showing their Architectures, Sizes, and Maturity. (``Est.'' = Estimated, ``B''= Billion Parameters)}
\centering
\renewcommand{\arraystretch}{1.1}
\resizebox{\textwidth}{!}{%
\begin{tabular}{l|c|l|l|c|c|c|c}
\textbf{LLM Name} & \textbf{Company} & \textbf{Model Size} & \textbf{Architecture} & \textbf{Release Date} & \textbf{Reasoning} & \textbf{Open Source} & \textbf{Code Instruct} \\
\hline
GPT-4o & OpenAI & 200 B \citep{howarth_gpt4_params} (est.) & Transformers (Decorder-only) & 6 August 2024 & No & No & No\\
\hline
GPT-5& OpenAI & \begin{tabular}[c]{@{}l@{}}125 B \citep{rbloggers_gpt5_parameters} --\\300 B \citep{lifearchitect_gpt5} (est.)\end{tabular} & Transformers (Decoder-only) & 7 August 2025 & Yes & No & No\\
\hline
Qwen2.5-Coder & \begin{tabular}[c]{@{}l@{}}Alibaba\\Cloud \end{tabular}& 14.7 B & \begin{tabular}[c]{@{}l@{}}Transformers with RoPE, \\ SwiGLU, RMSNorm\end{tabular} & November 2024 & No & Yes & Yes \\
\hline
Gemma-3 & Google & 12 B & \begin{tabular}[c]{@{}l@{}}Transformers (Decoder-only) with \\GQA, QK-Norm, and RMSNorm\end{tabular} & March 2025 & No & Yes & No\\
\hline
\begin{tabular}[c]{@{}l@{}}DeepSeek-V3.2\\(Non-Thinking)\end{tabular} & DeepSeek & 671 B & \begin{tabular}[c]{@{}l@{}}Multi-Head Latent Attention (MLA)\\and Transformer Hybrid\end{tabular} & 29 September 2025 & No & Yes & No\\
\hline
Codestral-2508 & Mistral AI & N/A & Transformers (Decoder-only) & July 2025 & No & No & Yes\\
\hline
LLaMA-3.1-8B & Meta AI & 8 B & Transformers (Decoder-only) & 23 July 2024 & No & Yes & No\\
\end{tabular}%
}
\label{tab:llm_architecture_comparison}
\end{table*}

\begin{table}[tb!]
\caption{\centering \textsc{Details of Tasks and Benchmarks}}
\vspace{-0.2em}
\renewcommand{\arraystretch}{1.1}
\resizebox{\columnwidth}{!}{%
\begin{tabular}{l|l|l|c|c}
\textbf{Dataset} & \textbf{Original Task} & \textbf{MuCoCo Task} & \textbf{Release Date} & \textbf{\# Qns.} \\
\hline
\begin{tabular}[c]{@{}l@{}}CodeMMLU\\(Code Completion)\end{tabular} & \begin{tabular}[c]{@{}l@{}}Multiple Choice\\Questions\end{tabular} & \begin{tabular}[c]{@{}l@{}}Multiple Choice\\Questions\end{tabular} & 9 Apr 2025 & 164 \\
\hline
HumanEval & Code Generation & \begin{tabular}[c]{@{}l@{}}Code Generation,\\ Input/Output Prediction\end{tabular} & 14 Jul 2021 & 164 \\
\hline
CruxEval & Input/Output Prediction & Input/Output Prediction  & 5 Jan 2024 & 800 \\
\hline
BigCodeBench  & Code Generation & Code Generation  & 1 Apr 2025 & 1140 \\
\end{tabular}%
}
\vspace{-0.75em}
\label{tab:dataset-summary}
\end{table}

\begin{table*}[tb!]
    \caption{\centering \approach's Effectiveness Across LLMs showing  inconsistency rate (Inc.) and Model Accuracy (Acc.).}
    \centering
    \renewcommand{\arraystretch}{1.15}
    \resizebox{\textwidth}{!}{
    \begin{tabular}{l|cc|cc|cc|cc|cc|cc|cc|cc}
        \textbf{Mutation} &
        \multicolumn{2}{c|}{\textbf{Qwen2.5}} &
        \multicolumn{2}{c|}{\textbf{Gemma-3}} &
        \multicolumn{2}{c|}{\textbf{DeepSeek-V3.2}} &
        \multicolumn{2}{c|}{\textbf{LLaMA-3.1}} &
        \multicolumn{2}{c|}{\textbf{GPT-5}} &
        \multicolumn{2}{c|}{\textbf{GPT-4o}} &
        \multicolumn{2}{c|}{\textbf{Codestral}} &
        \multicolumn{2}{c}{\textbf{All Models}} \\

      \textbf{Type}  & \textbf{Inc.} & \textbf{Acc.} &
          \textbf{Inc.} & \textbf{Acc.} &
          \textbf{Inc.} & \textbf{Acc.} &
          \textbf{Inc.} & \textbf{Acc.} &
          \textbf{Inc.} & \textbf{Acc.} &
          \textbf{Inc.} & \textbf{Acc.} &
          \textbf{Inc.} & \textbf{Acc.} &
          \textbf{Inc.} & \textbf{Acc.} \\
        \hline

        \textbf{Logical} &
        9.74 & 76.23 &
        15.52 & 66.13 &
        17.43 & 63.00 &
        18.02 & 49.88 &
        1.91 & 97.52 &
        19.48 & 64.25 &
        15.24 & 66.15 &
        13.91 (6259/45008) & 69.01 (31055/45000) \\

        \textbf{Syntactic} &
        6.72 & 77.91 &
        12.30 & 68.78 &
        14.89 & 65.80 &
        15.47 & 55.70 &
        1.51 & 98.46 &
        18.00 & 65.75 &
        12.24 & 68.71 &
        11.59 (2691/23220) & 71.58 (16612/23209) \\

        \textbf{Lexical} &
        16.17 & 67.63 &
        20.48 & 60.66 &
        16.50 & 69.24 &
        24.71 & 48.41 &
        3.91 & 94.10 &
        17.44 & 68.90 &
        14.75 & 70.96 &
        16.28 (12974/79707) & 68.48 (54425/79471) \\

        \hline
        \textbf{All (Avg.)} &
        12.73 & 71.86 &
        17.69 & 63.60 &
        16.53 & 66.80 &
        21.22 & 50.00 &
        2.92 & 95.84 &
        18.15 & 66.99 &
        14.51 & 69.14 &
        14.82 (21924/147935) & 69.13 (102092/147680) \\
    \end{tabular}
    }
    \label{tab:effectiveness-models}
\end{table*}

\begin{table*}[tb!]
    \caption{\centering 
    \approach's Effectiveness across tasks showing Inconsistency Rate (Inc.) and Accuracy (Acc.).}
    \centering
    \renewcommand{\arraystretch}{1.15}
    \resizebox{\textwidth}{!}{
    \begin{tabular}{l|cc|cc|cc|cc|cc}
        \textbf{Mutation} &
        \multicolumn{2}{c|}{\textbf{MCQ}} &
        \multicolumn{2}{c|}{\textbf{Input Pred.}} &
        \multicolumn{2}{c|}{\textbf{Output Pred.}} &
        \multicolumn{2}{c|}{\textbf{Code Gen.}} &
        \multicolumn{2}{c}{\textbf{All Tasks}} \\

        \textbf{Type} & \textbf{Inc.} & \textbf{Acc.} &
          \textbf{Inc.} & \textbf{Acc.} &
          \textbf{Inc.} & \textbf{Acc.} &
          \textbf{Inc.} & \textbf{Acc.} &
          \textbf{Inc.} & \textbf{Acc.} \\
        \hline

        \textbf{Logical} &
        38.50 & 46.48 &
        7.08 & 75.90 &
        18.52 & 64.16 &
        -- & -- &
        13.91 (6259/45008) & 69.01 (31055/45000) \\

        \textbf{Syntactic} &
        16.95 & 67.80 &
        5.90 & 80.53 &
        16.88 & 62.89 &
        -- & -- &
        11.59 (2691/23220) & 71.58 (16612/23209) \\

        \textbf{Lexical} &
        10.04 & 78.36 &
        6.62 & 80.59 &
        20.11 & 61.84 &
        27.15 & 57.66 &
        16.28 (12974/79707) & 68.48 (54425/79471) \\

        \hline
        \textbf{All (Avg.)} &
        22.35 & 64.04 &
        6.65 & 78.97 &
        18.99 & 62.82 &
        27.15 & 57.66 &
        14.82 (21924/147935) & 69.13 (102092/147680) \\
    \end{tabular}
    }
    \label{tab:effectiveness-tasks}
\end{table*}

\smallskip
\noindent
\textbf{3) Consistency Error Detection:}
\approach detects a \textit{consistency error} by comparing the LLM output and test suite correctness of the original task (step 1) versus the mutated task (step 2).  
A pair of original and mutated tasks is a \textit{consistency test case}.  
It then computes 
the \textit{consistency error rate} by aggregating the number of consistency test cases generated by \approach and the number of consistency errors triggered by \approach. 
This step is outlined in step 10 of \autoref{fig:workflow}
and it is repeated for all valid code mutations for each task and dataset.

\subsection{Semantic-preserving Mutations}

\approach 
provides 
11 mutation operators 
categorised 
into three (3) groups, namely lexical, syntactic and logical mutations. \autoref{sec:mutation-types} explains each mutation type/operator with examples. 

\smallskip 
\noindent
\textbf{a) Lexical Mutations}
refer to mutations involving renaming of variable names, function names, or Python strings.  
Lexical mutations include \texttt{random}, \texttt{sequential} and \texttt{literal format}.

\smallskip 
\noindent
\textbf{b) Syntactic Mutations} refers to semantic-preserving mutations that modify the program syntax with equivalent syntax, e.g., 
\texttt{for-to-enumerate} and \texttt{for-to-while}. 

\smallskip 
\noindent
\textbf{c) Logical Mutations}
modifies the program logic in a manner that preserves the original program behavior, e.g. decomposing a constant (5) into equivalent operations (4 + 1). 
This includes \texttt{DeMorgan mutation}, \texttt{boolean literal}, \texttt{commutative reorder} and \texttt{constant unfolding}. 


\subsection{\approach Test Oracle}
\label{sub:test-oracle}

\noindent
\approach employs two test oracles: (1) \textbf{Correctness Oracle}: The correctness oracle determines the correctness of an LLM output.
Correctness of LLM outputs are evaluated individually for each task. 
(2) \textbf{Consistency Oracle}: the performance of an LLM on a mutated dataset versus an original dataset. 
The test suite correctness of the mutated and original datasets are first evaluated using the correctness oracle.  
Next, each task in the original dataset is paired with its corresponding mutated version and their outputs and test suite correctness are compared. 
\approach's consistency oracle is applied to each pair of original and mutated tasks, and the resulting outcomes are aggregated to compute the consistency error rate (\textit{see \autoref{eq:inconsistency-score}}). 
Further details 
about the correctness and consistency oracles are provided in \autoref{sec:test-oracle-additional}.
We also report additional experiments investigating the direction of inconsistency and invalidity in \autoref{sec:inconsistency-direction}. 

\section{Experimental Settings}
\label{sec:experiment-setup}

\subsection{Research Questions}

We pose the following \textit{research questions} (RQs): 

\noindent
\textbf{RQ1 Effectiveness:} How effective is \approach in discovering code inconsistency in LLMs? 

\noindent
\textbf{RQ2 Baseline Comparison:} How does \approach compare to \turb? 


\noindent
\textbf{RQ3 Probing Study:} 
What is the impact of different mutators on \approach's effectiveness? 

\noindent
\recheck{We also report additional experiments in the Appendix, namely the effectiveness of \textit{multiple mutations }
(\autoref{sec:multiple-mutation-results}), 
 \textit{direction of inconsistency}
(\autoref{sec:inconsistency-direction}),    \textit{distance of inconsistency} (\autoref{sec:inconsistency-distance}) and  \textit{impact of model confidence} (\autoref{sec:model-confidence}). 
}


\subsection{Coding Tasks and Benchmarks}
\noindent
We evaluated  \approach 
using four benchmarks, namely BigCodeBench \citep{zhuo2025bigcodebench}, CodeMMLU \citep{nguyen2025codemmlu}, HumanEval \citep{chen2021evaluating} and CruxEval \citep{pmlr-v235-gu24c} (\textit{see }\autoref{tab:dataset-summary}). 
These benchmakrs provide the following coding tasks: Multiple Choice Questions (MCQ), Input Prediction, Output Prediction, Code Generation.
\autoref{sec:appendix_exp_setup} provides additional details about our experimental setup and task procedures.

\subsection{LLMs under Test}
\noindent
We evaluate \approach using seven (7) LLMs
(\textit{see }\autoref{tab:llm_architecture_comparison} for more details). 
We employ a diverse set of recent LLMs (2024-2025) including models with   
reasoning capabilities (GPT-5 \citep{openai_gpt5_docs}),  
varying model sizes (8-200 Billion) and architectures.  We also employ open-source/weight models (Llama \citep{grattafiori2024llama3herdmodels} and DeepSeek {\citep{deepseek_api_news_2025_09_29}}),  proprietary models (GPT-4o~\citep{openai_gpt4o_docs}),  and models that are trained on  code instruction fine-tuning (QwenCoder \citep{hui2024qwen25codertechnicalreport}).

\subsection{Metrics and Measures}
\noindent
\approach computes 
\textit{consistency error rate} (aka Inconsistency/Inc./ErrRate) as follows:

\begin{equation}
\text{Inc.} = \frac{N_{\text{inconsistent}}}{N_{\text{test\_cases}}} \times 100
\label{eq:inconsistency-score}
\end{equation}

\noindent
$N_\text{inconsistent}$ refers to the number of consistency errors discovered by \approach by comparing the LLM outputs and test suite correctness of \approach's 
mutated dataset versus the original dataset.
$N_\text{test\_cases}$ is the number of test cases 
(mutated and original programs pairs) 
generated 
by \approach. 

\subsection{Baseline Settings}
\noindent
We compare \approach  to the \turb~\citep{10989005} benchmark as a baseline for testing code inconsistencies (\textit{see}\autoref{sec:results} (\textbf{RQ2)}).
Since \turb 
uses a parametrised program templates for instantiating programs, 100 of these parameters are first created using the code base provided by the \turb authors 
\recheck{and a} pre-defined seed (\texttt{1234}).

To ensure the validity of the generated parameters, they are inserted into their corresponding templates and evaluated against the template's test suite.
Parameters that successfuly pass all the test cases are then stored.
During testing, these parameters are retrieved 
to recreate the \turb 
instances.
There are 60 templates in the \turb 
benchmark and 100 sets of parameters were generated for each template. 
In total,  we generated 5612 valid sets of parameters using \turb.
We note that the distribution of valid parameters across the templates is uneven.

\begin{table*}[tb!]
    \caption{\centering 
   \approach's Effectiveness across benchmarks showing 
  Inconsistency Rate (Inc.) and Accuracy (Acc.).}
    \centering
    \renewcommand{\arraystretch}{1.15}
    \resizebox{\textwidth}{!}{
    \begin{tabular}{l|cc|cc|cc|cc|cc}
        \textbf{Mutation} &
        \multicolumn{2}{c|}{\textbf{CodeMMLU}} &
        \multicolumn{2}{c|}{\textbf{HumanEval}} &
        \multicolumn{2}{c|}{\textbf{CruxEval}} &
        \multicolumn{2}{c|}{\textbf{BigCodeBench}} &
        \multicolumn{2}{c}{\textbf{All Datasets}} \\

        \textbf{Type} & \textbf{Inc.} & \textbf{Acc.} &
          \textbf{Inc.} & \textbf{Acc.} &
          \textbf{Inc.} & \textbf{Acc.} &
          \textbf{Inc.} & \textbf{Acc.} &
          \textbf{Inc.} & \textbf{Acc.} \\
        \hline

        \textbf{Logical} &
        38.50 & 46.48 &
        9.95 & 73.26 &
        21.17 & 60.57 &
        -- & -- &
        13.91 (6259/45008) & 69.01 (31055/45000) \\

        \textbf{Syntactic} &
        16.95 & 67.80 &
        8.35 & 74.94 &
        16.08 & 66.74 &
        -- & -- &
        11.59 (2691/23220) & \textbf{71.58 (16612/23209)} \\

        \textbf{Lexical} &
        10.04 & 78.36 &
        10.89 & 74.55 &
        17.42 & 66.41 &
        27.26 & 56.95 &
        16.28 (12974/79707) & 68.48 (54425/79471) \\

        \hline
        \textbf{All} &
        22.35 & 64.04 &
        10.09 & 74.10 &
        18.05 & 65.11 &
        27.26 & 56.95 &
        14.82 (21924/147935) & 69.13 (102092/147680) \\
    \end{tabular}
    }
    \label{tab:effectiveness-datasets}
\end{table*}

\section{Results}
\label{sec:results}
\noindent
\textbf{RQ1 Effectiveness:}
We examine the test effectiveness of \approach using all models, datasets, tasks and mutation operators. 
\autoref{tab:effectiveness-models}, \autoref{tab:effectiveness-tasks},  and \autoref{tab:effectiveness-datasets} highlights \approach's effectiveness across all tested models, tasks, and datasets, respectively. 

We found that \textit{\approach is effective in exposing inconsistencies in Code LLMs: \recheck{About one in seven (14.82\%) test cases generated by \approach exposed consistency errors across all settings}. } 
\approach exposed \recheck{thousands} of consistency errors ($\approx$\recheck{22K errors out of 148K test cases}),  despite a relatively high model accuracy \recheck{(69.13\% on average)}.  
 \autoref{tab:effectiveness-models} also shows that the consistency error rate exposed by \approach ranges widely across models.
In particular,  \texttt{GPT-5} \citep{openai_gpt5_docs} had the least consistency error rate \recheck{(2.92\%),} while \texttt{LLaMA} \citep{hf_meta_llama_3_1_8b} and \texttt{GPT-4o} \citep{openai_gpt4o_docs} produced the most consistency error rate with \recheck{21.22\% and 18.15\%} consistency error rate, respectively.  
We attribute the performance of \texttt{GPT-5} \citep{openai_gpt5_docs} to its reasoning attribute. We also attribute the relatively average consistency error rate for code-instruct LLMs (\texttt{CodeStral} \citep{mistral_codestral_25_08_news} and \texttt{Qwen2.5} \citep{hf_qwen2_5_coder_14b}) to the impact of code domain adaptation. 

\autoref{tab:effectiveness-tasks} and \autoref{tab:effectiveness-datasets} show that consistency errors are most prevalent for code generation tasks and \texttt{BigCodeBench} \citep{zhuo2025bigcodebench} dataset due to the higher complexity of the task/dataset.  
We observed lower consistency error rate (\recheck{6.65\%}) for input prediction tasks and \texttt{HumanEval} \citep{chen2021evaluating} datasets due to the simplicity of the problems.  
As an example,  \texttt{HumanEval} \citep{chen2021evaluating} programs are significantly simpler than those in \texttt{BigCodeBench} \citep{zhuo2025bigcodebench}.  
Overall, these results demonstrate that \approach effectively exposes coding inconsistency in LLMs and consistency errors are prevalent across LLMs. 

\begin{result}
\recheck{
\approach is effective in exposing consistency errors in Code LLMs:
14.82\% of its generated inputs 
triggered consistency errors. 
}
\end{result}

\smallskip \noindent
\textbf{RQ2 Baseline Comparison:}
This experiment compares the effectiveness of \approach to 
a closely-related robustness benchmark called 
\turb~\citep{10989005}.  
\turb provides manually-crafted, parameterised program templates,  such that parameters can be instantiated to form multiple instances of the same program.  
\autoref{tab:mucoco-turbulence-examples} 
shows a consistency error revealed by \approach even though \turb's instances are consistent.  
We compare pairs of parameterized instances of \turb versus executing \approach on \turb instances. 
\autoref{tab:turbulence-comparison} presents our findings.

We observed that \textit{\approach 
reveals more consistency errors than \turb.}
 \approach has a (\recheck{22.46\%}) higher consistency error rate than \turb \recheck{(33.8\% vs.  41.39\%)}. 
Specifically,  \autoref{tab:turbulence-comparison} shows that \approach reveals five times \recheck{(5X)} more consistency errors \recheck{(1,781,288 vs.  279,241)} than \turb. 
In addition,  \approach generates significantly larger test suites than \turb. 
\approach generates \recheck{(4X)} more consistency test cases than \turb \recheck{(4,302,713 vs.  826,084).}  
These results demonstrate the test generation effectiveness of \approach versus \turb.

Qualitative analysis shows that \textit{\approach discovers a fundamentally different type of consistency errors in comparison to \turb}.  
While \turb discovers inconsistencies across different parameterized instances of the same program,  \approach discovers inconsistencies across semantic-preserving lexical,  syntactic or logical variants of a program.  
Consider the \turb program \recheck{Q18 (\autoref{tab:mucoco-turbulence-examples}). } \approach's \texttt{DeMorgan} mutation on instances of \recheck{Q18} (column 3) results in the mutated program (column 4).  
\approach reveals a consistency error (\xmark) via a logical mutation of the \texttt{if} condition of the \texttt{return} statement \recheck{(line 3)}: 

{\centering \footnotesize
\texttt{if i \% \$0 == 0 or i \% \$1 == 0]} \quad $\rightarrow$ \\
\texttt{if \textbf{not} (\textbf{not} i \% \$0 == 0 and (\textbf{not} i \% \$1 == 0))]}
}

\recheck{
In this example (\recheck{Q18}),  \approach detects a consistency error (\xmark) since \approach's logical mutation results in a correct GPT-4o output for the first instance (``{\color{blue}-672}$=$-672''  in column 6),  but an incorrect output for the logically modified instance (``{\color{red}486} $\neq$ 698'' in column 8).  
Meanwhile,  performing \textit{only} parameter instantiation for this query (\recheck{Q18}), as envisioned in \turb, does \textit{not} reveal a consistency error (\cmark). 
Specifically,  different parameter instantiations (\$0- \$3) in  
instance one (column 5) versus instance two (column 7) are consistently incorrect (\cmark) -- since their expected output consistently fails to match the expected output for both instances-- 
``{\color{red}-1114}  $\neq$ -672'' (column 6) and ``{\color{red}486} $\neq$ 698'' (column 8), respectively.  } 
Since this is an output prediction task, the expected output and actual GPT-4o answer are compared. 
As such, the model outputs are consistent for \turb instances, according to \autoref{tab:inconsistency-attribution} (\textit{last row}). 
This example demonstrates that \approach complements \turb: In comparison to \turb, \approach discovers different classes of coding inconsistency, while exposing more inconsistencies.

\begin{result}
 \recheck{ \approach is up to five times (5X) more effective than \turb in test generation.
It discover 22.46\% more consistency errors than \turb. 
}
\end{result}

\begin{table*}[tb!]
\caption{ \centering \textsc{Turbulence versus MuCoCo Instances}. \autoref{tab:mucoco-turbulence-examples-full} in Appendix provides additional examples. } 
\label{tab:mucoco-turbulence-examples}
\setlength{\tabcolsep}{4pt}
\begin{adjustbox}{max width=\textwidth}
\renewcommand{\arraystretch}{1.1}
\begin{tabular}{c|c|l|l|l|l|c|c|c|c}
\toprule
\textbf{ID} &
\begin{tabular}[c]{@{}c@{}}\textbf{Mutation}\\\textbf{Type}\end{tabular} &
\multicolumn{1}{c|}{\textbf{Turbulence Template}} &
\multicolumn{1}{c|}{\textbf{Turbulence Template + MuCoCo Mutation}} &
\multicolumn{1}{c|}{\begin{tabular}[c]{@{}c@{}}\textbf{Instance 1}\\\textbf{Parameters}\end{tabular}} &
\multicolumn{1}{c|}{\begin{tabular}[c]{@{}c@{}}\textbf{Instance 1}\\\textbf{LLM Output}\end{tabular}} &
\multicolumn{1}{c|}{\begin{tabular}[c]{@{}c@{}}\textbf{Instance 2}\\\textbf{Parameters}\end{tabular}} &
\multicolumn{1}{c|}{\begin{tabular}[c]{@{}c@{}}\textbf{Instance 2}\\\textbf{LLM Output}\end{tabular}} &\begin{tabular}[c]{@{}l@{}}\textbf{Turbulence}\\\textbf{Inc.}\end{tabular} &
\begin{tabular}[c]{@{}l@{}}\textbf{MuCoCo}\\\textbf{Inc.}\end{tabular} \\
\midrule

\textbf{Q18} & DeMorgan &
\begin{minipage}{0.55\linewidth}
\vspace{-6pt}
\begin{lstlisting}[language=Python, escapechar=!]
  def sum_ints_div_by_either_nums(l: List[int]) -> int:
    l = l[$2:$3 + 1]
    return !\textbf{sum([i for i in l if i \% \$0 == 0 or i \% \$1 == 0])}!
\end{lstlisting}
\vspace{-4pt}
\end{minipage} &
\begin{minipage}{0.55\linewidth}
\vspace{-6pt}
\begin{lstlisting}[language=Python, escapechar=!]
  def sum_ints_div_by_either_nums(l: List[int]) -> int:
    l = l[$2:$3 + 1]
    return !\textbf{sum([i for i in l if not (not i \% \$0 == 0 and (not i \% \$1 == 0))])}!
\end{lstlisting}
\vspace{-4pt}
\end{minipage} & 
\begin{tabular}[c]{@{}l@{}}
\textbf{Template Params:}\\
\texttt{  \$0: -4}\\
\texttt{  \$1: 4}\\
\texttt{  \$2: 1}\\
\texttt{  \$3: 6}\\ \\
\textbf{Program Input:}\\ 
\texttt{\lbrack212, -451, -512, 337,}\\
\texttt{486, -442, -160, -422 \rbrack} \\\\
\end{tabular} & 
\begin{tabular}[c]{@{}l@{}}
\textbf{Expected Output:}\\
\texttt{-672} \\\\
\textbf{Turbulence Output:}\\
\texttt{\color{red}-1114}\\\\
\textbf{MuCoCo Output:}\\
\texttt{\color{blue}-672}\\\\
\end{tabular} &
\begin{tabular}[c]{@{}l@{}}
\textbf{Template Params:}\\
\texttt{  \$0: -6}\\
\texttt{  \$1: 4}\\
\texttt{  \$2: 2}\\
\texttt{  \$3: 4}\\ \\
\textbf{Program Input:}\\ 
\texttt{\lbrack 337, -512, 212,}\\
\texttt{-451, 486, -422\rbrack}\\ \\
\end{tabular} & 
\begin{tabular}[c]{@{}l@{}}
\textbf{Expected Output:}\\
\texttt{698}\\\\
\textbf{Turbulence Output:}\\
\texttt{\color{red}486}\\\\
\textbf{MuCoCo Output:}\\
\texttt{\color{red}486}\\\\
\end{tabular} &
\cmark & \xmark \\
\midrule

\end{tabular}
\end{adjustbox}
\end{table*}

\begin{table*}[tb!]
    \caption{\centering Effectiveness of \approach versus \turb using the \turb dataset.  Best performance per (sub)category is marked in \textbf{bold} text.  (``\#'' = Number of ,  "errs'' = Consistency Errors,  ``tests'' = Test Cases,  ``ErrRate'' = Error Rate, ``Impr. '' = Improvement)}  %
    \centering
 \renewcommand{\arraystretch}{1.15}
    \resizebox{\textwidth}{!}{
    \begin{tabular}{l|ccc|ccc|ccc|ccc}
        \multirow{2}{*}{\textbf{Approach}}   &  \multicolumn{3}{c|}{\textbf{Code Generation}} & \multicolumn{3}{c|}{\textbf{Input Prediction}} & \multicolumn{3}{c|}{\textbf{Output Prediction}} & \multicolumn{3}{c}{\textbf{All Tasks}} \\
        & \textbf{\#errs} & \textbf{\#tests} & \textbf{ErrRate} & \textbf{\#errs} & \textbf{\#tests} & \textbf{ErrRate} & \textbf{\#errs} & \textbf{\#tests} & \textbf{ErrRate} & \textbf{\#errs} & \textbf{\#tests} & \textbf{ErrRate} \\
        \hline
 		\textbf{\turb} & 20359 & 272888 & \textbf{7.46\%}	& 54809 & 276598 & \textbf{19.82\%} & 204073 & 276598 & 73.78\% & 279241 & 826084 & 33.8\%  \\  
 		 \textbf{\approach}
 		 & \textbf{31458} & \textbf{542627} & 5.80\% &	\textbf{363593} & \textbf{1882518} & 19.31\% & \textbf{1386284} & \textbf{1877568} &  \textbf{73.83\%} & \textbf{1781288} & \textbf{4302713} & \textbf{41.39\%} \\
 		 \hline
    \textbf{\% Impr.} & {54.52\%} & {98.85\%} & {-22.25\%} & {563.38\%}  & {580.60\%} &  {-2.57\%} & {579.31\%} & {578.81\%} & {0.07\%} & {537.90\%} & {420.86\%} & {22.46\%} \\
    \end{tabular}
    }
    \label{tab:turbulence-comparison}
\end{table*}

\smallskip
\noindent
\textbf{RQ3 Probing Study:}
We inspect the impact of \approach's 
\textit{mutators} on its effectiveness.

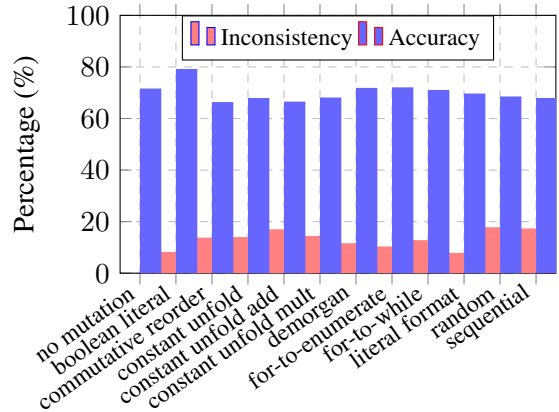
\begin{figure}[tb!]
\caption{\centering \textsc{Consistency error rate and accuracy by mutation operator.}}
\begin{tikzpicture}
\vspace{-0.25em}
\begin{axis}[
    width=0.95\columnwidth,
    height=5cm,
    ybar,
    bar width=8pt,
    ymin=0, ymax=100,
    ylabel={Percentage (\%)},
    symbolic x coords={
        no mutation,
        boolean literal,
        commutative reorder,
        constant unfold,
        constant unfold add,
        constant unfold mult,
        demorgan,
        for\--to\--enumerate,
        for\--to\--while,
        literal format,
        random,
        sequential
    },
    xtick=data,
    xticklabels={
        no mutation,
        boolean literal,
        commutative reorder,
        constant unfold,
        constant unfold add,
        constant unfold mult,
        demorgan,
        for\--to\--enumerate,
        for\--to\--while,
        literal format,
        random,
        sequential
    },
    xticklabel style={rotate=35, anchor=east, font=\footnotesize},
    enlarge x limits=0.05,
    grid=major,
    grid style=dashed,
    ymajorgrids=true,
    bar shift=-4pt,
    legend style={
        at={(0.5,1)},
        anchor=north,
        legend columns=-1,
        font=\small
    }
]

\addplot+[ybar, fill=red!50, draw=none] coordinates {
    (no mutation, 0.00)
    (boolean literal, 8.23)
    (commutative reorder, 13.78)
    (constant unfold, 14.01)
    (constant unfold add, 17.03)
    (constant unfold mult, 14.41)
    (demorgan, 11.63)
    (for\--to\--enumerate, 10.36)
    (for\--to\--while, 12.81)
    (literal format, 7.87)
    (random, 17.83)
    (sequential, 17.35)
};

\addplot+[ybar, fill=blue!60, draw=none, bar shift=4pt] coordinates {
    (no mutation, 71.61)
    (boolean literal, 79.20)
    (commutative reorder, 66.37)
    (constant unfold, 67.95)
    (constant unfold add, 66.55)
    (constant unfold mult, 68.14)
    (demorgan, 71.87)
    (for\--to\--enumerate, 72.08)
    (for\--to\--while, 71.07)
    (literal format, 69.69)
    (random, 68.54)
    (sequential, 67.96)
};

\legend{Inconsistency, Accuracy}
\end{axis}
\end{tikzpicture}
\vspace{-0.75em}
\label{fig:mutation-inc-acc}
\end{figure}

\autoref{fig:mutation-inc-acc} shows  that \approach's lexical mutations (random and sequential renaming) and logical mutations (constant unfolding and DeMorgan) induce the most consistency errors across all settings.  
\autoref{tab:effectiveness-models},  \autoref{tab:effectiveness-tasks} and \autoref{tab:effectiveness-datasets} also show that \approach's logical and lexical mutation operators induce the most consistency errors, \recheck{13.91\% and 16.28\%}, respectively.  
We attribute the higher consistency error rate for logical mutations to their higher code complexity.  Meanwhile,  we attribute the higher  consistency error rate for lexical mutations to LLM's inherent natural language dependence, in particular,  dependence on meaningful variable names and method names. 

\begin{result}
\recheck{
Logical and lexical mutations induce a higher consistency error rate than syntactic mutations.  
}
\end{result}

Inspecting the mutation operators that impact model accuracy, we observed that boolean literal improves model accuracy versus the original program without mutation (``no\_mutation''  in \autoref{fig:mutation-inc-acc}).  We also found that syntactic mutations (\texttt{for-to-enumerate})
and logical mutation (DeMorgan) also slightly improve model accuracy.  Most mutations improve or preserve model accuracy, except logical mutations (commutative reordering and constant unfolding).  We attribute this to the complexity of logical mutations which makes it more challenging for LLMs.  
Overall,  these results imply that LLMs are less robust to method/variable renaming and changes in program logic.  

\begin{result}
\recheck{
Logical mutations induce a lower model accuracy than the original program or other mutation types (lexical and syntactic mutations).   
}
\end{result}

\section{Additional Experiments}
\label{sec:add-experiments}
In additional experiments, we find that inconsistencies were often induced by \approach's mutated query rather than the original query 
(\autoref{sec:inconsistency-direction}).
We also find that most inconsistencies (55.08\%) discovered by \approach are correctness-based inconsistencies (\autoref{sec:inconsistency-count-type}).
Moreover,  we measure the 
inconsistency distance across models, we observe that open-source/-weight models have a greater inconsistency distance than proprietary models 
(\autoref{sec:inconsistency-distance}).  
\camcheck{In addition,  we examined two mitigation approaches for alleviating inconsistency in Code LLMs. 
The \textit{first} mitigation approach explores the impact of model confidence on inconsistency using open-source/-weight models. 
We found 
that increasing the confidence threshold 
reduces inconsistencies in LLMs %
from \revise{8.03\%} to \revise{1.89\%} 
(\autoref{sec:model-confidence}). 
Our \textit{second} mitigation approach adapts mitigation techniques used in Kontest \citep{rajan-etal-2024-knowledge}, where we found that ensemble techniques reduce inconsistency rates from 14.82\% to 2.94\% (\autoref{sec:model-ensemble}).}
Finally, we inspect the impact of complex mutation operators (i.e., multiple (two) mutations) on discovering inconsistency in Code LLMs. 
Our findings (\autoref{sec:multiple-mutation-results}) show that multiple mutations led to a \revise{5.81\% } increase in inconsistency in comparison to atomic mutations.

\section{Related Works}
\label{sec:related-works}

\noindent
\textbf{LLM Robustness and Consistency Testing:}
Honarvar et al. \citep{10989005} presented the \turb benchmark as a novel template-based approach for evaluating LLM correctness and robustness.
LLM robustness is evaluated by comparing the differences in outputs within the same neighbourhood \citep{10989005}. 
\approach differs from \turb by assessing LLM inconsistency through code mutation on existing benchmarks, and conducting one-to-one comparison of results of mutated and original tasks.
\approach also includes tasks like MCQ, input prediction and output prediction tasks.
The KONTEST framework proposed by Rajan et al. \citep{rajan-etal-2024-knowledge} exposes knowledge gaps in LLMs through an automatic testing framework, utilizing a knowledge graph approach to construct test cases.
On the other hand, \approach utilizes rule-based mutation of existing benchmarks for testing.
KONTEST also differs in scope with \approach, it is focused on general knowledge testing \citep{rajan-etal-2024-knowledge}, while \approach is aimed at exposing inconsistencies in Code LLMs.

\smallskip
\noindent
\textbf{Code LLM Benchmarks and Evaluation Methods:}
The HumanEval benchmark \citep{chen2021evaluating} is a well-established dataset used for evaluating LLM's correctness through code generation tasks. 
BigCodeBench by Zhuo et al. \citep{zhuo2025bigcodebench} presents a series of diverse Python code generation tasks aimed at challenging LLMs with a range of function calls.
CodeMMLU \citep{nguyen2025codemmlu} introduces a comprehensive dataset of MCQ-type questions, testing LLMs on a variety of code-related tasks such as code repair, execution reasoning and code completion. 
CruxEval \citep{pmlr-v235-gu24c} is a benchmark focused on testing LLM code knowledge through input prediction and output prediction tasks.
The aforementioned benchmark datasets are primarly focused on evaluating the correctness of LLMs and are static by nature.
Meanwhile, \approach automatically transforms an existing coding dataset to generate test suites for evaluating code consistency. 
Orvalho et al. \citep{orvalho2026largelanguagemodelsrobust} proposed a code mutation framework for testing LLM robustness.
However, their approach exposed code inconsistencies through manual, expert analysis of the LLM's reasoning \citep{orvalho2026largelanguagemodelsrobust}, which differs from \approach's automatic testing approach. 

\smallskip
\noindent
\textbf{Dynamic Testing Approaches:}  
Guan et al. \citep{guan2026benchmarkusefuldynamicbenchmarking} proposed a dynamic benchmarking framework to keep static benchmarks relevant over time.
Similar to \approach, their dynamic testing framework utilizes code mutation to transform static benchmarks including \texttt{VarNormI}, \texttt{VarNormII}, \texttt{For-to-While} and \texttt{ConstUnfold}~\citep{guan2026benchmarkusefuldynamicbenchmarking}.
However, the dynamic testing framework is targeted at evaluating LLM's correctness \citep{guan2026benchmarkusefuldynamicbenchmarking}, which differs from \approach which is targeted at exposing code inconsistencies.
Additionaly, \approach supports a total of 11 mutation variations, offering a greater diversity than the dynamic benchmark framework.
\approach also broadens the LLM evaluation methodology by including MCQ and code generation tasks, while the dynamic testing framework tests with code execution and code translation \citep{guan2026benchmarkusefuldynamicbenchmarking}.

\section{Conclusion} 
\label{sec:conclusion}
In this work,  we propose an automated testing approach for assessing the consistency property of LLMs for coding tasks.  Our technique (\approach) employs mutation analysis and metamorphic properties to discover code consistency errors in LLMs.  It provides 11 semantic-preserving mutation operators spanning lexical,  logical and syntactic mutations. 
We evaluate \approach using seven LLMs,  four benchmarks and four datasets.  Results show that \approach is effective in consistency testing and code inconsistency is prevalent in state-of-the-art LLMs.  We also compare \approach to a state-of-the-art baseline (\turb) and demonstrate its superiority and complementarity.  Our work motivates the need to automatically assess LLMs for other properties, beyond correctness, in particular, consistency. 


\section{Limitations}
\label{sec:threats}

\noindent
\textbf{Construct Validity:} The main threat to construct validity of this work is the prompting of LLMs and processing of LLM outputs.  For each dataset,  we have used the best prompting option out of the three settings (zero-shot, one-shot, few-shot), based on the availability of examples or lack thereof.  In addition,  to ensure that LLM outputs are processed correctly, we have used system prompts to ensure models provide responses in the expected format.  We have also tested this with the canonical solution (before mutation) to ensure correctness.   We have also used open-source/-weight models (Llama and DeepSeek) with frozen weights  to mitigate and reduce non-determinism.  We have set temperature to zero for all models and reasoning to minimal for GPT-5 to mitigate randomness.  Finally, we mitigate the effect of model updates by limiting our experiments to a short time and tracking update news.   

\smallskip
\noindent 
\textbf{Internal Validity:} This refers to the risk that our implementation of \approach performs the intended code consistency testing. To mitigate this, we have conducted manual and automated tests, code review and manual inspection of mutation operators,  sampled experiments and results.  In  addition, we have conducted probing studies (\autoref{sec:results} \textbf{RQ3} to examine the correctness of \approach. Finally, we have reported additional ablation and sensitivity studies (\autoref{sec:add-experiments} and Appendix) to examine the stability/robustness of \approach.  

\smallskip
\noindent
\textbf{External Validity:} The main threat to external validity is the generalizability of our approach and findings. 
To mitigate this, we have employed several state-of-the-art models, tasks and datasets with varying sizes, maturity and complexity.  We acknowledge that our approach has been applied and evaluated using only Python programs, thus, it  may not directly apply to other programming languages (PLs).  However,  we note that several mutations in our approach apply to most high-level programming languages, especially our lexical and logical mutations.  We acknowledge that syntactic mutations are PL dependent and may require redesign for other languages.  

\smallskip
\noindent
\textbf{Prompt Configuration: } Our experiments are conducted using the strongest available prompt configuration for each benchmark, selecting among zero-shot, one-shot, and few-shot prompting \citep{brown2020language}. 
However, this choice is also constrained by the design of the benchmark datasets. 
In particular, CruxEval \citep{pmlr-v235-gu24c} only supports zero-shot prompting, while BigCodeBench \citep{zhuo2025bigcodebench} supports up to one-shot prompting.
Although few-shot prompting is widely considered as the superior prompt technique, such configurations are not uniformly available across benchmarks. 
As a result, performance comparisons across datasets may partially reflect differences in allowable prompt configurations rather than model capability alone.

\smallskip
\noindent
\textbf{Model Randomness:} LLMs exhibit inherent stochasticity in their generation process. 
To mitigate this, all models except GPT-5 \citep{openai_gpt5_docs} are evaluated with a temperature setting of zero (0). 
GPT-5 API does not offer the option to set model temperature hence we set its reasoning mode to \texttt{minimal} \citep{openai_latest_model_guide}. 
Despite these measures, nondeterminism cannot be fully eliminated, and our results may therefore be affected by some stochastic variations in model behavior

\section{Ethical Considerations}
This section outlines the ethical considerations associated with our study.

\smallskip
\noindent
\textbf{Datasets.}
We exclusively use established benchmark datasets obtained from their official releases on HuggingFace.
These datasets have been made publicly available, widely used by the research community 
~\citep{zhuo2025bigcodebench, pmlr-v235-gu24c, nguyen2025codemmlu,chen2021evaluating}.

\smallskip
\noindent
\textbf{Methodology and Model Usage.}
Pre-trained LLMs by the official companies were used to evaluate \approach. 
We hereby acknowledge that LLMs may reflect biases present in their training data or modeling assumptions.
Additionally, we limit our experiments to inference on pre-trained models, thereby avoiding the significant environmental costs (such as energy and water consumption) incurred from model training.

\section*{Acknowledgements}
\camcheck{This project is supported by the Ministry of Education, Singapore, under its MOE AcRF Tier 1 (SUTD Kickstarter Initiative (SKI) SKI 2021 07 01). We would also like to express sincere gratitude to the DSO-AISG Incentive Award Programme for their financial support for Chua Jin Chou, which made the travel and presentation of this research possible. We appreciate Sarang Nambiar, Dhruv Pradhan and Srinivasan M for providing invaluable feedback on the previous drafts of this paper. We also acknowledge and appreciate the anonymous reviewers for their insightful comments on improving our work.}

\bibliography{custom}

\appendix

\section{\approach Test Oracle}
\label{sec:test-oracle-additional}
This section contains additional details for \approach's correctness and consistency oracles.

\noindent
\textbf{Correctness Oracle}: The correctness oracle determines the correctness of an LLM output.
Correctness of LLM outputs are evaluated individually for each task. 
For code generation, the program generated by the LLM is tested against the 
test suite provided by the respective benchmarks.
For input and output prediction tasks, the LLM’s response is evaluated for equivalence against the expected output specified in the benchmark's test suite.
Lastly,  inconsistency in 
MCQ task  is determined by checking whether the LLM’s output (i.e.,  selected option) matches the canonical output in the benchmark.
A correctness score is not assigned when \approach's code mutation 
is not possible, i.e.,  the original benchmark task cannot be mutated (e.g., due to the absence of a valid \texttt{for}-loop).
Formally,  \approach's \textit{correctness} oracle is defined as: 

\begin{equation}
\mathcal{O}_{\text{corr}}(A_i) =
\begin{cases}
\texttt{Pass}, & \text{if } A_i \text{ is correct,} \\[4pt]
\texttt{AssertionError}, & \text{if } A_i \text{ is incorrect,} \\[4pt]
\texttt{Invalid Answer}, & \text{otherwise.}
\end{cases}
\end{equation}

\noindent 
where $A_i$ and $\mathcal{O}_{\text{corr}}$ refers to the LLM answer and the test correctness test oracle, respectively.
\texttt{Invalid Answer} occurs when the LLM failed to return valid code (for code generation tasks), or if the LLM failed to return the answer within the given time budget.

\noindent
\textbf{Consistency Oracle}: To determine (in)consistency,  \approach's consistency oracle compares the performance of an LLM on a mutated dataset versus an original dataset. 
The test suite correctness of the mutated and original datasets are first evaluated using the correctness oracle.  
Next, each task in the original dataset is paired with its corresponding mutated version and their outputs and test suite correctness are compared. 
Inconsistencies between LLM outputs are categorised into the following:

\noindent (1) \textbf{Correctness-based inconsistency}: occurs when with one of the LLM output is correct and the other is incorrect.

\noindent(2) \textbf{Incorrectness-based inconsistency}: occurs when both LLM outputs are incorrect, and have differing degrees of incorrectness. 
For code generation tasks, a typical test suite comprises of multiple test cases. 
The two incorrect LLM outputs are evaluated with the test suites and the sets of test cases passed are compared.
If there is a difference in the two sets, an incorrectness-based inconsistency has occurred.
Otherwise, the LLM outputs are consistent.
For other tasks, the LLM outputs are compared directly, and an inconsistency is recorded if they differ.

\noindent(3) \textbf{Invalidity-based inconsistency}: occurs when only one of the LLM output is invalid, while the other can either be correct or incorrect.
Finally, no inconsistency occurs when both LLM outputs are correct or if both are invalid.

\autoref{tab:inconsistency-attribution} offers an overview of their inconsistency attribution for each pair of mutation types.
\approach's consistency oracle is applied to each pair of original and mutated tasks, and the resulting outcomes are aggregated to compute the consistency error rate (\textit{see \autoref{eq:inconsistency-score}}). 
Additionally, additional experiments were conducted to investigate the direction of inconsistency and invalidity.
These findings are elaborated in \autoref{sec:inconsistency-direction}.

\section{Types of Mutations Supported by \approach}
\label{sec:mutation-types}
\noindent
This section provides a more detailed description of the types of mutations supported by \approach.

\begin{table*}[tb!]
\centering
\caption{
\centering
\textsc{MuCoCo Mutation Types and Model Results. 
(\cmark = consistency,  \xmark = inconsistency (aka consistency error), OG = LLM Answer for Original Question, MU = LLM Answer for Mutated Question)}}
\label{tab:mucoco-examples}
\begin{adjustbox}{max width=\textwidth}
\renewcommand{\arraystretch}{1.1}
\setlength{\tabcolsep}{3pt}
\begin{tabular}{c|l|l|c|c|c|c|c|c|c|c}
\toprule
\textbf{Mutation Type} & \multicolumn{1}{c|}{\textbf{Original Question}} & \multicolumn{1}{c|}{\textbf{Mutated Question}} & \begin{tabular}[c]{@{}l@{}}\textbf{Dataset}\\\textbf{(Canon. Ans.)} \end{tabular} & 
\textbf{\begin{tabular}[c]{@{}c@{}} GPT-4o \\ (OG) \\ (MU) \end{tabular}} & 
\textbf{\begin{tabular}[c]{@{}c@{}} GPT-5 \\ (OG) \\ (MU) \end{tabular}} & 
\textbf{\begin{tabular}[c]{@{}c@{}} Qwen \\ (OG) \\ (MU) \end{tabular}} & 
\textbf{\begin{tabular}[c]{@{}c@{}} Llama \\ (OG) \\ (MU) \end{tabular}} & 
\textbf{\begin{tabular}[c]{@{}c@{}} Gemma \\ (OG) \\ (MU) \end{tabular}} & 
\textbf{\begin{tabular}[c]{@{}c@{}} DeepSeek \\ (OG) \\ (MU) \end{tabular}} & 
\textbf{\begin{tabular}[c]{@{}c@{}} Codestral \\ (OG) \\ (MU) \end{tabular}} \\
\midrule

\textbf{Random} &
\begin{minipage}{0.35\linewidth}
\vspace{-6pt}
\begin{lstlisting}[language=Python, escapechar=@]
  def f(text, n):
    if n < 0 or len(text) <= n:
      return text
    result = text[0 : n]
    i = len(result) - 1
    while i >= 0:
      if result[i] != text[i]:
        break
      i -= 1
    return text[0 : i + 1]
\end{lstlisting}
\vspace{-4pt}
\end{minipage} &
\begin{minipage}{0.45\linewidth}
\vspace{-6pt}
\begin{lstlisting}[language=Python, escapechar=@]
  def dOdtZZSuisMB(TYesQRBm, mXDbZtJMZFV):
    if mXDbZtJMZFV < 0 or len(TYesQRBm) <= mXDbZtJMZFV:
      return TYesQRBm
    result = TYesQRBm[0:mXDbZtJMZFV]
    i = len(result) - 1
    while i >= 0:
      if result[i] != TYesQRBm[i]:
        break
      i -= 1
    return TYesQRBm[0:i + 1]
\end{lstlisting}
\vspace{-4pt}
\end{minipage} &
\begin{tabular}[c]{@{}c@{}}CruxEval\_789 \\ ('br')\end{tabular} & 
\begin{tabular}[c]{@{}c@{}} \xmark \\ ('br') \\ (['bR']) \end{tabular} &
\cmark & 
\begin{tabular}[c]{@{}c@{}} \xmark \\ ('br') \\ (['bR']) \end{tabular} &
\cmark & 
\begin{tabular}[c]{@{}c@{}} \xmark \\ ('br') \\ (['bR']) \end{tabular} &
\begin{tabular}[c]{@{}c@{}} \xmark \\ ('br') \\ (['bR']) \end{tabular} &
\begin{tabular}[c]{@{}c@{}} \xmark \\ ('br') \\ (['bR']) \end{tabular} \\


\textbf{Sequential} &
\begin{minipage}{0.45\linewidth}
\vspace{-4pt}
\begin{lstlisting}[language=Python, escapechar=!]
  def f(text):
    return not text.isdecimal()
\end{lstlisting}
\vspace{-4pt}
\end{minipage} &
\begin{minipage}{0.45\linewidth}
\vspace{-4pt}
\begin{lstlisting}[language=Python, escapechar=!]
  def generic_function1(var1):
    return not var1.isdecimal()
\end{lstlisting}
\vspace{-4pt}
\end{minipage} &
\begin{tabular}[c]{@{}c@{}}CruxEval\_518 \\ (True)\end{tabular} & 
\cmark & 
\cmark & 
\cmark & 
\begin{tabular}[c]{@{}c@{}} \xmark \\ (True) \\ (False) \end{tabular} &
\cmark & 
\begin{tabular}[c]{@{}c@{}} \xmark \\ (True) \\ (False) \end{tabular} &
\cmark\\


\textbf{Literal Format} &
\begin{minipage}{0.5\linewidth}
\vspace{-4pt}
\begin{lstlisting}[language=Python, escapechar=@]
  def f(name):
    new_name =''
    name = name[::-1]
    for i in range(len(name)):
      n = name[i]
      if n !='.' and  new_name.count('.')<2:
        new_name=n+new_name
      else:
        break
    return new_name
\end{lstlisting}
\vspace{-4pt}
\end{minipage} &
\begin{minipage}{0.52\linewidth}
\vspace{-4pt}
\begin{lstlisting}[language=Python, escapechar=@]
  def f(name):
    new_name = "" @\textcolor{codegreen}{\texttt{\# ' > "}}@
    name = name[::-1]
    for i in range(len(name)):
      n = name[i]
      if n != "." and new_name.count(".") < 2: @\textcolor{codegreen}{\texttt{\# ' > "}}@
        new_name = n + new_name
      else:
        break
    return new_name
  return a
\end{lstlisting}
\vspace{-4pt}
\end{minipage} &
\begin{tabular}[c]{@{}c@{}}CruxEval\_273 \\ ("NET") \end{tabular} & 
\cmark & 
\cmark & 
\cmark & 
\begin{tabular}[c]{@{}c@{}} \xmark \\ ("NET") \\ (".NET") \end{tabular} &
\begin{tabular}[c]{@{}c@{}} \xmark \\ ("NET") \\ ("TEN") \end{tabular} &
\cmark & 
\cmark\\

\hline

\textbf{For2Enumerate} &
\begin{minipage}{0.4\linewidth}
\vspace{-4pt}
\begin{lstlisting}[language=Python, escapechar=!]
def f(text):
    x = 0
    if text.islower():
      for c in text:
        if int(c) in list(range(90)):
          x+=1
    return x
\end{lstlisting}
\vspace{-4pt}
\end{minipage} &
\begin{minipage}{0.45\linewidth}
\vspace{-4pt}
\begin{lstlisting}[language=Python, escapechar=!]
  def f(text):
    x = 0
    if text.islower():
      for loop_var0, c in enumerate(text):
        if int(c) in list(range(90)):
          x += 1
    return x
\end{lstlisting}
\vspace{-4pt}
\end{minipage} &
\begin{tabular}[c]{@{}c@{}}CruxEval\_560 \\ (0)\end{tabular} & 
\cmark & 
\cmark & 
\begin{tabular}[c]{@{}c@{}} \xmark \\ (0) \\ (9) \end{tabular} &
\cmark & 
\cmark & 
\cmark & 
\cmark \\


\textbf{For2While} &
\begin{minipage}{0.42 \linewidth}
\vspace{-4pt}
\begin{lstlisting}[language=Python, escapechar=@]
  def f(string):
    l = list(string)
    for i in reversed(range(len(l))):
      if l[i] != ' ':
        break
      l.pop(i)
    return ''.join(l)
\end{lstlisting}
\vspace{-4pt}
\end{minipage} &
\begin{minipage}{0.48\linewidth}
\vspace{-4pt}
\begin{lstlisting}[language=Python, escapechar=@]
  def f(string):
    l = list(string)
    new_reversed_var4 = range(len(l))[::-1]
    length_var3 = len(new_reversed_var4)
    loop_var0 = 0
    while loop_var0 < length_var3:
      i = new_reversed_var4[loop_var0]
      if l[i] != ' ':
        break
      l.pop(i)
      loop_var0 += 1
    return ''.join(l)
\end{lstlisting}
\vspace{-4pt}
\end{minipage} &
\begin{tabular}[c]{@{}c@{}}CruxEval\_708 \\ ('    jcmfxv')\end{tabular} & 
\cmark & 
\cmark & 
\begin{tabular}[c]{@{}c@{}} \xmark \\ ('    jcmfxv') \\ ('jcmfxv     ') \end{tabular} &
\cmark & 
\cmark & 
\begin{tabular}[c]{@{}c@{}} \xmark \\ ('    jcmfxv') \\ ('jcmfxv') \end{tabular} &
\begin{tabular}[c]{@{}c@{}} \xmark \\ ('    jcmfxv') \\ ('jcmfxv') \end{tabular} \\
\hline

\textbf{DeMorgan} &
\begin{minipage}{0.41\linewidth}
\vspace{-4pt}
\begin{lstlisting}[language=Python, escapechar=@]
  def f(array, target):
    count, i = 0, 1
    for j in range(1, len(array)):
      if ((array[j] > array[j-1]) and (array[j] <= target)): 
        count += i
      elif array[j] <= array[j-1]: 
        i = 1
      else: 
        i += 1
    return count
\end{lstlisting}
\vspace{-4pt}
\end{minipage}  &
\begin{minipage}{0.48\linewidth}
\vspace{-4pt}
\begin{lstlisting}[language=Python, escapechar=@]
  def f(array, target):
    count, i = (0, 1)
    for j in range(1, len(array)):
      if not (not array[j] > array[j - 1] or not array[j] <= target):
        count += i
      elif array[j] <= array[j - 1]:
        i = 1
      else:
        i += 1
    return count
\end{lstlisting}
\vspace{-4pt}
\end{minipage} &
\begin{tabular}[c]{@{}c@{}}CruxEval\_223 \\ (1)\end{tabular} & 
\begin{tabular}[c]{@{}c@{}} \xmark \\ (1) \\ \verb|(```\n2\n```)| \end{tabular} &
\cmark & 
\cmark & 
\begin{tabular}[c]{@{}c@{}} \xmark \\ (1) \\ \verb|(2)| \end{tabular} &
\cmark & 
\begin{tabular}[c]{@{}c@{}} \xmark \\ (1) \\ \verb|(0)| \end{tabular} &
\begin{tabular}[c]{@{}c@{}} \xmark \\ (2) \\ \verb|(1)| \end{tabular} \\


\textbf{Boolean Literal} &
\begin{minipage}{0.34\linewidth}
\vspace{-4pt}
\begin{lstlisting}[language=Python, escapechar=!]
  def f(lst):
    lst.clear()
    for i in lst:
      if i == 3:
        return False
    else:
      return True
\end{lstlisting}
\vspace{-4pt}
\end{minipage} &
\begin{minipage}{0.36\linewidth}
\vspace{-4pt}
\begin{lstlisting}[language=Python, escapechar=!]
  def f(lst):
    lst.clear()
    for i in lst:
      if i == 3:
        return not True
    else:
      return not False
\end{lstlisting}
\vspace{-4pt}
\end{minipage} &
\begin{tabular}[c]{@{}c@{}}CruxEval\_97 \\ (\texttt{True})\end{tabular} & 
\cmark & 
\cmark & 
\cmark & 
\begin{tabular}[c]{@{}c@{}} \xmark \\ (True) \\ (False) \end{tabular} &
\cmark & 
\begin{tabular}[c]{@{}c@{}} \xmark \\ (True) \\ (False) \end{tabular} &
\begin{tabular}[c]{@{}c@{}} \xmark \\ (True) \\ (False) \end{tabular} \\


\textbf{Commutative Reorder} &
\begin{minipage}{0.5\linewidth}
\vspace{-4pt}
\begin{lstlisting}[language=Python, escapechar=!]
  def f(text, letter):
    if letter in text:
      start = text.index(letter)
      return text[start + 1:] + text[:start + 1]
    return text
\end{lstlisting}
\vspace{-4pt}
\end{minipage} &
\begin{minipage}{0.5\linewidth}
\vspace{-4pt}
\begin{lstlisting}[language=Python, escapechar=!]
  def f(text, letter):
    if letter in text:
      start = text.index(letter)
      return text[1 + start:] + text[:1 + start]
    return text
\end{lstlisting}
\vspace{-4pt}
\end{minipage} &
\begin{tabular}[c]{@{}c@{}} CruxEval\_786 \\ (\texttt{"kefp719"}) \end{tabular} & 
\begin{tabular}[c]{@{}c@{}} \xmark \\ ("kefp719") \\ ("19kefp7") \end{tabular} &
\cmark & 
\begin{tabular}[c]{@{}c@{}} \xmark \\ ("kefp719") \\ ("9kefp71") \end{tabular} &
\cmark & 
\begin{tabular}[c]{@{}c@{}} \xmark \\ ("9kefp71") \\ ("kefp719") \end{tabular} &
\begin{tabular}[c]{@{}c@{}} \xmark \\ ("kefp719") \\ ("efp719k") \end{tabular} &
\cmark \\


\textbf{Constant Unfolding} &
\begin{minipage}{0.4\linewidth}
\vspace{-4pt}
\begin{lstlisting}[language=Python, escapechar=!]
  def f(n):
    p = ''
    if n % !\textbf{2}! == 1:
      p+='sn'
    else:
      return n * n
    for x in range(1, n+1):
      if x % !\textbf{2}! == 0:
        p+= 'to'
      else:
        p+= 'ts'
    return p
\end{lstlisting}
\vspace{-4pt}
\end{minipage} &
\begin{minipage}{0.5\linewidth}
\vspace{-4pt}
\begin{lstlisting}[language=Python, escapechar=!]
  def f(n):
    p = ''
    if n % !\textbf{(1 + 1)}! == 1:
      p += 'sn'
    else:
      return n * n
    for x in range(1, n+1):
      if x % !\textbf{(2 * 1)}! == 0:
        p += 'to'
      else:
        p += 'ts'
    return p
\end{lstlisting}
\vspace{-4pt}
\end{minipage} &
\begin{tabular}[c]{@{}c@{}}CruxEval\_506 \\ (\texttt{"snts"}) \end{tabular}  & 
\begin{tabular}[c]{@{}c@{}} \xmark \\ ("snts") \\ ("sn\_ts") \end{tabular} &
\cmark & 
\begin{tabular}[c]{@{}c@{}} \xmark \\ ("snts") \\ ("sn") \end{tabular} &
\begin{tabular}[c]{@{}c@{}} \xmark \\ ("snts") \\ ("sn") \end{tabular} &
\cmark & 
\begin{tabular}[c]{@{}c@{}} \xmark \\ ("sn") \\ ("snts") \end{tabular} &
\cmark\\


\textbf{\begin{tabular}[c]{@{}c@{}}Constant Unfolding\\(Add) \end{tabular}} &
\begin{minipage}{0.4\linewidth}
\vspace{-4pt}
\begin{lstlisting}[language=Python, escapechar=!]
  def f(nums):
    for i in range(len(nums)):
      if nums[i] % 3 == 0:
        nums.append(nums[i])
    return nums
\end{lstlisting}
\vspace{-4pt}
\end{minipage} &
\begin{minipage}{0.45\linewidth}
\vspace{-4pt}
\begin{lstlisting}[language=Python, escapechar=!]
  def f(nums):
    for i in range(len(nums)):
      if nums[i] % (1 + 2) == 0 + 0:
        nums.append(nums[i])
    return nums
\end{lstlisting}
\vspace{-4pt}
\end{minipage} &
\begin{tabular}[c]{@{}c@{}}CruxEval\_226 \\ (\texttt{[1,3,3]}) \end{tabular} &
\cmark & 
\cmark & 
\cmark &
\begin{tabular}[c]{@{}c@{}} \xmark \\ ([1,3,3]) \\ ([1,3,1,3]) \end{tabular} &
\begin{tabular}[c]{@{}c@{}} \xmark \\ ([1,3,3]) \\ ([1,3]) \end{tabular} &
\begin{tabular}[c]{@{}c@{}} \xmark \\ ([1,3,3]) \\ ([1,3,1,3]) \end{tabular} &
\begin{tabular}[c]{@{}c@{}} \xmark \\ ([1,3,3]) \\ ([1,3,1,3]) \end{tabular} \\

\textbf{\begin{tabular}[c]{@{}c@{}}Constant Unfolding\\(Multi) \end{tabular}} &
\begin{minipage}{0.42\linewidth}
\vspace{-4pt}
\begin{lstlisting}[language=Python, escapechar=!]
  def f(lists):
    lists[!\textbf{1}!].clear()
    lists[!\textbf{2}!] += lists[!\textbf{1}!]
    return lists[!\textbf{0}!]
\end{lstlisting}
\vspace{-4pt}
\end{minipage} &
\begin{minipage}{0.45\linewidth}
\vspace{-4pt}
\begin{lstlisting}[language=Python, escapechar=!]
  def f(lists):
    lists[!\textbf{1 * 1}!].clear()
    lists[!\textbf{2 * 1}!] += lists[!\textbf{1 * 1}!]
    return lists[!\textbf{0 * 1}!]
\end{lstlisting}
\vspace{-4pt}
\end{minipage} &
\begin{tabular}[c]{@{}c@{}}CruxEval\_564 \\ (\texttt{[395,666,7,4]})\end{tabular} & 
\cmark & 
\cmark & 
\cmark &
\begin{tabular}[c]{@{}c@{}} \xmark \\ ([395,666,7,4,666]) \\ ([395,666,7,4]) \end{tabular} &
\cmark &
\begin{tabular}[c]{@{}c@{}} \xmark \\ ([395,666,7,4]) \\ ([4223,111]) \end{tabular} &
\begin{tabular}[c]{@{}c@{}} \xmark \\ (395) \\ ([395,666,7,4]) \end{tabular} \\
\bottomrule
\end{tabular}
\end{adjustbox}
\end{table*}

\subsubsection{Lexical Mutations}
Lexical mutations refer to mutations involving renaming variable names, function names, or Python strings.  
These mutations modify the natural language names used in code without modifying the program logic. 
Code inconsistencies exposed via lexical mutations suggest that the LLM may not fully understand the program logic and relies more on meaningful names for program comprehension. 
Guan et al. \citep{guan2026benchmarkusefuldynamicbenchmarking} proposed similar variable renaming mutations \textbf{\textit{VarNormI}} and \textbf{\textit{VarNormII}} in a dynamic benchmarking framework and Orvalho et al. \citep{orvalho2026largelanguagemodelsrobust} also proposed \textbf{Variable Renaming} mutation, which is similar to Random mutation.
The mutations under this category includes the following:

\noindent
\textbf{Random Mutation} replaces the function/variable names with random strings (\textit{see Table }\autoref{tab:mucoco-examples} row one).

\noindent
\textbf{Sequential Mutation } is similar to random mutation,  but it replaces function/variable names with generic names.
By default, the function name in the function definition is renamed to \verb|generic_function1| and first variable names are replaced with \verb|var1| and so on (\textit{see Table }\autoref{tab:mucoco-examples} row two).

\noindent
\textbf{Literal Format}
replaces string single quotation marks (') with double quotation (") and vice versa (\textit{see Table }\autoref{tab:mucoco-examples} row three).

\subsubsection{Syntactic Mutations}
It refers to semantic-preserving mutations that modify the 
program syntax with equivalent syntax. 
Both Guan et al. \citep{guan2026benchmarkusefuldynamicbenchmarking} and Orvalho et al. \citep{orvalho2026largelanguagemodelsrobust} proposed for-to-while mutation in their respective works.
The mutations under this category include:

\noindent
\textbf{for-to-enumerate} replaces all \texttt{for} loop iterators with \texttt{enumerate} iterators 
(\textit{see Table }\autoref{tab:mucoco-examples} row four).

\noindent
\textbf{for-to-while Mutation}
replaces all \texttt{for} loops with \texttt{while} loops.  
(\textit{see Table }\autoref{tab:mucoco-examples} row five).

\subsubsection{Logical Mutations}
Logical mutations 
modifies the program logic in a manner that preserves the original program behavior, e.g., by decomposing a constant (5) into equivalent operations (4 + 1). 
\textbf{Constant Unfolding} and \textbf{Condition Augmentation} are similar program mutations proposed by Guan et al. \citep{guan2026benchmarkusefuldynamicbenchmarking}.
The mutations under this category include:

\noindent
\textbf{DeMorgan Mutation:}
De Morgan's Law refers to a pair of rules that convert logical operations with conjunction operators into disjunction equivalents. For instance, 
\begin{equation}
\texttt{A and B} \rightarrow \texttt{not A or not B}
\end{equation}
\begin{equation}
\texttt{A or B} \rightarrow \texttt{not A and not B}
\end{equation}

\autoref{tab:mucoco-examples} (row six) illustrates how \approach applies the DeMorgan's law on programs containing  logical 
statements. 

\noindent
\textbf{Boolean Literal}
transforms \texttt{True} and \texttt{False}, into semantically-equivalent expression using negation (\textit{see Table }\autoref{tab:mucoco-examples} row seven).

\noindent
\textbf{Commutative Reorder:} The order of variables in 
commutative expressions (such as summation and multiplication) are reversed to form a new expression (\textit{see Table }\autoref{tab:mucoco-examples} row eight).

\noindent
\textbf{Constant Unfolding:}
Integers are decomposed into two factors and expressed using either multiplication or addition expressions (\textit{see Table }\autoref{tab:mucoco-examples} row nine).

\noindent
\textbf{Constant Unfolding (Add):} 
It substitutes all valid constants 
with equivalent addition expressions (\textit{see Table }\autoref{tab:mucoco-examples} row 10).

\noindent
\textbf{Constant Unfolding (Multiplication)} replaces 
constants with equivalent multiplication expressions (\textit{see Table }\autoref{tab:mucoco-examples} row 11).

\section{Task-specific Experimental Setup}
\label{sec:appendix_exp_setup}
\noindent
This section provides additional details on \approach's experimental setup and testing procedures for Multiple Choice Questions, Input Prediction, Output Prediction and Code Generation tasks.

\subsubsection{Multiple Choice Questions (MCQ)}
Using the CodeMMLU benchmark \citep{nguyen2025codemmlu},    
the LLM is tasked with selecting the correct completion of an incomplete program 
when given 
a description of the program including the incomplete code,  along with several completion options.
Four completion options are provided to the  LLM  
(\texttt{A,B,C,D}) and the LLM has to choose the option that completes the program. 
\subsubsection{Input Prediction}
The LLM is tasked with 
predict whether a provided input will result in given output when the provide program is executed.  We provide the LLM with a canonical program,  a program description and an input-output pair.  
The allowed LLM response is either 
\texttt{True} or \texttt{False}.  
We employ CruxEval \citep{pmlr-v235-gu24c} and HumanEval \citep{chen2021evaluating}. 
\recheck{We employ the HumanEval dataset for this task by splitting its test cases into input - output pairs, which are then used for input prediction.}
Hence,  the expected LLM output is always \texttt{True}. 

\subsubsection{Output Prediction}
The LLM is tasked with predicting the correct program output 
when given the input, the canonical program and a program description.
CruxEval \citep{pmlr-v235-gu24c} and HumanEval \citep{chen2021evaluating} benchmarks are used for this task.
Test pairs for HumanEval benchmark were generated using the same approach for Input Prediction.
The Guan et al. \citep{guan2026benchmarkusefuldynamicbenchmarking} and Orvalho et al. \citep{orvalho2026largelanguagemodelsrobust} utilized output prediction on CruxEval for LLM testing in their respective works.

\subsubsection{Code Generation}
The LLM is tasked with generating a program when given a program description and a snippet of the program.  
The snippet only contains the function definition and input variable names.
The LLM must complete the program such that it matches the program description and in accordance to the given function and variable names.
BigCodeBench \citep{zhuo2025bigcodebench} and HumanEval \citep{chen2021evaluating} are used for this task.

\section{Inconsistency Errors per Error Type}
\label{sec:inconsistency-count-type}
\autoref{tab:inconsistency-type-distribution} reports the distribution of consistency errors per error type.  
We found that \textit{correctness-based inconsistency} accounts for the majority of inconsistencies found by \approach (55.08\%), this is followed by \textit{in}correctness-based inconsistency (44.05\%).
Invalidity-based inconsistencies contributes the least number of inconsistencies, accounting for only 0.42\% of consistency errors.
\autoref{tab:inconsistency-type-distribution} shows that logical mutations causes the most number of inconsistencies (59.18\%).
We attiribute this to the complexity of code generation tasks which only have lexical mutations due to the lack of program snippets. 
Meanwhile, logical and syntactic mutations account for 28.55\% and 12.27\% of the discovered inconsistencies.
It should be noted that syntactic mutations have the least number of mutations (\texttt{for-to-while} and \texttt{for-to-enumerate}). These results suggest that LLMs are more robust to mutations in (Python) programming syntax than lexical or logical changes in programs.  


\begin{table}[tb!]
\centering
\caption{\centering Distribution of inconsistency types across mutation categories.}
\renewcommand{\arraystretch}{1.15}
\resizebox{\columnwidth}{!}{%
\setlength{\tabcolsep}{6pt}
\begin{tabular}{l | l r r}
\toprule
\textbf{Inconsistency Type} & \textbf{Mutation Type} & \textbf{Count} & \textbf{Percentage} \\
\midrule
\multirow{3}{*}{\begin{tabular}[c]{@{}l@{}}Incorrectness-based\\Inconsistency\end{tabular}}
 & Logical   & 2401 & 10.95\% \\
 & Syntactic & 1213 & 5.53\%  \\
 & Lexical   & 6143 & 28.02\% \\
\midrule
\multirow{3}{*}{\begin{tabular}[c]{@{}l@{}}Correctness-based\\Inconsistency\end{tabular}}
 & Logical   & 3825 & 17.45\% \\
 & Syntactic & 1460 & 6.66\%  \\
 & Lexical   & 6790 & 30.97\% \\
\midrule
\multirow{3}{*}{\begin{tabular}[c]{@{}l@{}}Invalidity-based\\Inconsistency\end{tabular}}
 & Logical   & 33   & 0.15\%  \\
 & Syntactic & 18   & 0.08\%  \\
 & Lexical   & 41   & 0.19\%  \\
\midrule
\textbf{All} 
& \textbf{All} & \textbf{21924} & \textbf{100.00\%} \\
\bottomrule
\end{tabular}
}
\label{tab:inconsistency-type-distribution}
\end{table}

\section{Inconsistency Direction in \approach}
\label{sec:inconsistency-direction}


We examine the direction of inconsistency in order to attribute the cause of inconsistency to \approach's mutation or the original input.  As an example, a correctness-based inconsistency arising from the LLM's output being correct on the original task and incorrect on the mutated counterpart has an incorrectness direction caused by mutation.
Additionally, we also account for the invalidity direction to determine if applying mutation operators led to more invalid LLM outcomes. \autoref{tab:directional-attribution} provides the directional attribution for incorrectness and invalidity for all possible outcome pairs.

\autoref{tab:directional-inconsistency}  highlights our findings. 
Across all three mutation categories, we find that mutated tasks are consistently the primary contributor to inconsistency.  
In addition,  most incorrectness are caused by mutated tasks. 
On the other hand, we find that original dataset is the primary invalidity direction for lexical and syntactic mutations, while mutations are the invalidity direction for logical mutations.
However, original tasks contributes more to the invalidity direction. 
Overall,  these results show that the consistency errors in Code LLMs, reported in this work,  are caused by \approach's mutations, rather than the original inputs.  

\begin{table}[tb!]
\centering
\caption{\centering Directional inconsistency attribution based on original (OG) and mutated (Mut) task outcomes.}
\renewcommand{\arraystretch}{1.15}
\setlength{\tabcolsep}{6pt}
\resizebox{\columnwidth}{!}{%
\begin{tabular}{llcc}
\toprule
\textbf{OG Task} & \textbf{Mut Task} & \textbf{Incorr Dir} & \textbf{Inv Dir} \\
\midrule
Correct   & Incorrect & Mut        & -- \\
Correct   & Correct   & --         & -- \\
Correct   & Invalid   & --         & Mut \\
\midrule
Incorrect & Incorrect & OG + Mut   & -- \\
Incorrect & Correct   & OG         & -- \\
Incorrect & Invalid   & OG         & Mut \\
\midrule
Invalid   & Incorrect & Mut        & OG \\
Invalid   & Correct   & --         & OG \\
Invalid   & Invalid   & --         & OG + Mut \\
\bottomrule
\end{tabular}
}
\label{tab:directional-attribution}
\end{table}

\begin{table}[tb!]
\centering
\caption{\centering Directional inconsistency statistics across mutation categories.}
\renewcommand{\arraystretch}{1.15}
\resizebox{\columnwidth}{!}{%
\begin{tabular}{l|cc|cc}
\toprule
\textbf{Category} & \textbf{\# Mut} & \textbf{\# OG} & \textbf{Inc. Dir} & \textbf{Inv. Dir} \\
\midrule
Lexical Incorrect Dir     & 25{,}022 & 24{,}401 & Mut & -- \\
Logical Incorrect Dir     & 13{,}939 & 12{,}562 & Mut & -- \\
Syntactic Incorrect Dir   & 6{,}597  & 6{,}452  & Mut & -- \\
\midrule
Lexical Invalid Dir       & 268  & 431  & -- & OG \\
Logical Invalid Dir       & 26   & 19   & -- & Mut \\
Syntactic Invalid Dir     & 11   & 14   & -- & OG \\
\midrule
Aggregated Incorrect Dir & 45{,}558 & 43{,}415 & Mut & -- \\
Aggregated Invalid Dir   & 305 & 464 & -- & OG \\
\bottomrule
\end{tabular}
}
\label{tab:directional-inconsistency}
\end{table}

\section{Inconsistency Distance in \approach}
\label{sec:inconsistency-distance}

We propose an additional metric for quantifying the degree of inconsistency between LLM outputs. This involves measuring the correctness differences across test suites.
This metric applies only to code generation tasks, which include multiple test cases within test suites for evaluating output correctness.
Unlike binary correctness judgments,  employing the test suites proffers a more fine-grained assessment of consistency by determining the specific set of test cases that pass or fail. This metric leverages the test suite sensitivity of each model outcome to quantify the degree of consistency errors. 
By conducting a pair-wise comparison between the outcome of each test case of the LLM output on the original and mutated task, we compute the inconsistency distance between the outputs.

\noindent
As an example,  consider a test suite containing three unique test cases for testing correctness on the generated LLM programs.
For simplicity, the test cases will be referred to as first, second, and third. Let us assume that,  for the original task, the LLM-generated program passed the first and third test cases, while failing the second one.
Meanwhile,  for the mutated task, the LLM-generated program passed the second and third test cases and failed the first.
The correctness of the two solutions differ for two test cases (first and second) and only aligned for the third function.
Since there are a total of three test cases within the test suite, this leads to an inconsistency distance of $2/3$.

\noindent
Concretely, the inconsistency distance (Inc Dist) for a test suite is shown in \autoref{eq:inconsistency-distance}. 

\begin{equation}
\label{eq:inconsistency-distance}
 \text{Inc Dist} = \frac{\text{\# of inconsistent test cases}}{\text{\# of test cases}}
\end{equation}

\noindent
We can then aggregate the inconsistency distance (Agg Inc Dist) of each task to determine the model with the highest inconsistency distance using \autoref{eq:agg-inconsistency-distance}.

\begin{equation}
\label{eq:agg-inconsistency-distance}
  \text{Agg Inc Dist} =\frac{\sum \text{Inconsistency Distance}}{\text{Total Questions}}
\end{equation}

\noindent 
\autoref{tab:inc-dist-comparison} highlights the consistency distance results.
Our findings determined that proprietary models have a smaller inconsistency distance when compared to open-source/-weight models, with a $0.084$ difference.
This suggests that proprietary models are more consistent than open-source/-weight models even when we consider more fine-grained metrics.
We also find that GPT-5 \citep{openai_gpt5_docs} is the most consistent model for code generation, with an aggregated inconsistency distance of $0.094$.
On the other hand, Llama-3.1-8b \citep{grattafiori2024llama3herdmodels} is the most inconsistent model with an aggregated inconsistent distance of $0.235$.

\section{Implementation Details and Platform}
\label{sec:exp-implementation}
\noindent
\approach utilizes \texttt{AST} Python package \citep{python_ast_library} for code mutation processes.
Qwen \citep{hf_qwen2_5_coder_14b}, Llama \citep{hf_meta_llama_3_1_8b} and Gemma \citep{hf_gemma_3_12b_it} model weights were downloaded using the HuggingFace Transformers Python package \citep{huggingface_transformers_docs}. 
The experiments on these models were then conducted on Google Colab \citep{google_colab}, using A100 Tensor Core GPU with 80GB RAM.
Experiments on GPT-5 \citep{openai_gpt5_docs}, GPT-4o \citep{openai_gpt4o_docs}, Codestral \citep{mistral_codestral_25_08_news} and Deepseek \citep{deepseek_api_news_2025_09_29} models were conducted using the official APIs provided by their website.

\begin{table*}[tb!]
\centering
\caption{\centering 
\approach's Scalability to multiple (2) mutations 
vs. atomic mutations. 
("Inc." = Inconsistency, "Acc." =  Accuracy)}
\renewcommand{\arraystretch}{1.1}
\resizebox{\textwidth}{!}{%
\begin{tabular}{l|l|cc|cc|cc|cc|cc|cc}

\multirow{3}{*}{\rotatebox[origin=c]{90}{\shortstack[c]{\textbf{Mutatn.}\\ \textbf{Level}}}}& 
&
\multicolumn{4}{c|}{\textbf{Input Prediction}} &
\multicolumn{4}{c|}{\textbf{Output Prediction}} &
\multicolumn{2}{c|}{\textbf{MCQ}} &
\\

& \multicolumn{1}{c|}{\textbf{Mutation Operation}}  
&
\multicolumn{2}{c|}{\textbf{HumanEval}} &
\multicolumn{2}{c|}{\textbf{CruxEval}} &
\multicolumn{2}{c|}{\textbf{HumanEval}} &
\multicolumn{2}{c|}{\textbf{CruxEval}} &
\multicolumn{2}{c|}{\textbf{CodeMMLU}} &
\multicolumn{2}{c}{\textbf{All Tasks \& Datasets}} \\
& \multicolumn{1}{c|}{(type)} & \textbf{Inc.} & \textbf{Acc.} 
 & \textbf{Inc.} & \textbf{Acc.} 
 & \textbf{Inc.} & \textbf{Acc.} 
 & \textbf{Inc.} & \textbf{Acc.} 
 & \textbf{Inc.} & \textbf{Acc.}
 & \textbf{Inc.} & \textbf{Acc.} \\
\hline
 & No Mutation & N/A & 75.38 & N/A & 80.75 & N/A & 59.10 & N/A & 53.82 & N/A & 85.71 & N/A & 67.92 (2528/3722) \\
\hline 
\multirow{3}{*}{\rotatebox[origin=c]{90}{\textbf{Atomic}}} & Random (lexical) & 9.75 & 75.28 & 13.25 & 77.25 & 23.42 & 58.49 & 33.12 & 52.12 & 7.58 & 87.12 & 19.53 (701/3590) & 64.63 (2491/3854) \\
& Constant Unfold (logical) & 10.20 & 72.80 & 7.73 & 78.45 & 23.00 & 59.20 & 41.99 & 35.56 & 40.98 & 44.26 & 18.80 (256/1362) & 61.90 (918/1483) \\
& For2while (syntactic) & 10.49 & 73.66 & 15.24 & 77.14 & 22.84 & 56.17 & 32.38 & 48.57 & 22.03 & 69.49 & 19.48 (312/1602) & 62.85 (1081/1720) \\
\hline

\multirow{3}{*}{\rotatebox[origin=c]{90}{\shortstack[c]{\textbf{Second}\\ \textbf{-order}}}} & Constant Unfold + Random & 12.02 & 72.53 & 14.36 & 76.24 & 28.51 & 69.04 & 44.20 & 43.09 & 39.34 & 50.82 & 22.71 (290/1277) & 65.69 (919/1399) \\
& For2While + Random & 13.17 & 72.37 & 17.78 & 69.65 & 32.10 & 70.58 & 48.25 & 52.06 & 30.51 & 61.02 & 26.72 (428/1602) & 65.81 (1130/1717) \\
& For2While + Constant Unfold & 13.03 & 77.39 & 13.68 & 71.58 & 33.72 & 64.37 & 53.68 & 38.30 & 50.00 & 46.67 & 26.12 (186/712) & 65.24 (503/771) \\
\Xhline{1pt}
& Atomic (All) & 10.05 & \textbf{74.26} & 12.96 & \textbf{77.39} & 23.17 & 58.10 & 34.18 & \textbf{48.96} & 19.05 & \textbf{72.62} & 19.36 (1269/6554) & 63.62 (4490/7057) \\
& Second-order (All) & \textbf{12.70} & 73.51 & \textbf{16.07} & 71.99 & \textbf{31.10} & \textbf{68.65} & \textbf{47.88} & 47.12 & \textbf{38.00} & 54.00 & 25.17 (904/3591) & 65.65 (2552/3887) \\
\end{tabular}%
}
\label{tab:second-order-mutation}
\end{table*}

\begin{table*}[t]
\centering
\caption{\centering \textsc{Turbulence vs. MuCoCo Instances Extended Examples}}
\label{tab:mucoco-turbulence-examples-full}
\setlength{\tabcolsep}{4pt}
\begin{adjustbox}{max width=\textwidth}
\renewcommand{\arraystretch}{1.1}
\begin{tabular}{c|c|l|l|l|l|c|c|c|c}
\toprule
\textbf{ID} &
\begin{tabular}[c]{@{}c@{}}\textbf{Mutation}\\\textbf{Type}\end{tabular} &
\multicolumn{1}{c|}{\textbf{Turbulence Template}} &
\multicolumn{1}{c|}{\textbf{Turbulence Template + MuCoCo Mutation}} &
\multicolumn{1}{c|}{\begin{tabular}[c]{@{}c@{}}\textbf{Instance 1}\\\textbf{Parameters}\end{tabular}} &
\multicolumn{1}{c|}{\begin{tabular}[c]{@{}c@{}}\textbf{Instance 1}\\\textbf{LLM Output}\end{tabular}} &
\multicolumn{1}{c|}{\begin{tabular}[c]{@{}c@{}}\textbf{Instance 2}\\\textbf{Parameters}\end{tabular}} &
\multicolumn{1}{c|}{\begin{tabular}[c]{@{}c@{}}\textbf{Instance 2}\\\textbf{LLM Output}\end{tabular}} &\begin{tabular}[c]{@{}l@{}}\textbf{Turbulence}\\\textbf{Inc.}\end{tabular} &
\begin{tabular}[c]{@{}l@{}}\textbf{MuCoCo}\\\textbf{Inc.}\end{tabular} \\
\midrule

\textbf{Q33} & For2While &
\begin{minipage}{0.55\linewidth}
\vspace{-6pt}
\begin{lstlisting}[language=Python, escapechar=!]
 def return_vowels(s: str) -> List[str]:
   vowels = ['a','e','i','o','u','A','E','I','O','U']
   result = []
   sliced_s = s[$0:$1]
   if not sliced_s:
     return result
   !\textbf{for char in sliced\_s:}!
     if char in vowels and '$2' < char <= '$3':
       result.append(char)
   return result
\end{lstlisting}
\vspace{-4pt}
\end{minipage} &
\begin{minipage}{0.55\linewidth}
\vspace{-6pt}
\begin{lstlisting}[language=Python, escapechar=!]
 def return_vowels(s: str) -> List[str]:
   vowels = ['a','e','i','o','u','A','E','I','O','U']
   result = []
   sliced_s = s[$0:$1]
   if not sliced_s:
     return result
   !\textbf{loop\_var0 = 0}!
   !\textbf{while loop\_var0 < len(sliced\_s):}!
     !\textbf{char = sliced\_s[loop\_var0]}!
     if char in vowels and '$2' < char <= '$3':
       result.append(char)
     !\textbf{loop\_var0 += 1}!
   return result
\end{lstlisting}
\vspace{-4pt}
\end{minipage} & 
\begin{tabular}[c]{@{}l@{}}
\textbf{Template Params:}\\
\texttt{  \$0: 22}\\
\texttt{  \$1: 44}\\
\texttt{  \$3: 'z'}\\
\texttt{  \$4: '\{\}'}\\ \\
\textbf{Program Input:}\\ 
\texttt{"31 w 014\%Xh>00a4"}\\ \\
\end{tabular} & 
\begin{tabular}[c]{@{}l@{}}
\textbf{Expected Output:}\\
\texttt{[]}\\\\
\textbf{Turbulence Output:}\\
\texttt{[]}\\\\
\textbf{MuCoCo Output:}\\
\texttt{[]}\\\\
\end{tabular} &
\begin{tabular}[c]{@{}l@{}}
\textbf{Template Params:}\\
\texttt{  \$0: 14}\\
\texttt{  \$1: 36}\\
\texttt{  \$3: 'l'}\\
\texttt{  \$4: 'm'}\\ \\
\textbf{Program Input:}\\ 
\texttt{"4 X4h\$<a04 g10"}\\ \\
\end{tabular} &
\begin{tabular}[c]{@{}l@{}}
\textbf{Expected Output:}\\
\texttt{[]}\\\\
\textbf{Turbulence Output:}\\
\texttt{[]}\\\\
\textbf{MuCoCo Output:}\\
\texttt{[]}\\\\
\end{tabular}
& \cmark & \cmark \\
\midrule

\textbf{Q39} & \begin{tabular}[c]{@{}c@{}}Constant \\ Unfolding\end{tabular} &
\begin{minipage}{0.55\linewidth}
\vspace{-6pt}
\begin{lstlisting}[language=Python, escapechar=!]
 def return_n_greatest_chars(s: str) -> List[str]:
   if len(s) < $0:
     return []
   if "$1" == "ascending":
     sorted_s = sorted(s)
     return !\textbf{sorted\_s[:\$0]}!
   else:
     sorted_s = sorted(s, reverse=True)
     return !\textbf{sorted\_s[:\$0]}!
\end{lstlisting}
\vspace{-4pt}
\end{minipage} &
\begin{minipage}{0.55\linewidth}
\vspace{-6pt}
\begin{lstlisting}[language=Python, escapechar=!]
 def return_n_greatest_chars(s: str) -> List[str]:
   if len(s) < $0:
     return []
   if "$1" == "ascending":
     sorted_s = sorted(s)
     return !\textbf{sorted\_s[:\$0*1]}!
   else:
     sorted_s = sorted(s, reverse=True)
     return !\textbf{sorted\_s[:\$0*1]}!
\end{lstlisting}
\vspace{-4pt}
\end{minipage} & 
\begin{tabular}[c]{@{}l@{}}
\textbf{Template Params:}\\
\texttt{  \$0: 2}\\
\texttt{  \$1: "descending"}\\ \\
\textbf{Program Input:}\\ 
\texttt{"oal.e\textasciicircum"}\\ \\
\end{tabular} & 
\begin{tabular}[c]{@{}l@{}}
\textbf{Expected Output:}\\
\texttt{['o','l']}\\\\
\textbf{Turbulence Output:}\\
\texttt{['\textasciicircum','o']}\\\\
\textbf{MuCoCo Output:}\\
\texttt{['\textasciicircum','o']}\\\\
\end{tabular} &
\begin{tabular}[c]{@{}l@{}}
\textbf{Template Params:}\\
\texttt{  \$0: 5}\\
\texttt{  \$1: descending}\\ \\
\textbf{Program Input:}\\ 
\texttt{"oal.e\textasciicircum|kmTEcdS@"}\\ \\
\end{tabular} & 
\begin{tabular}[c]{@{}l@{}}
\textbf{Expected Output:}\\
\texttt{['|','o','m','l','k']}\\\\
\textbf{Turbulence Output:}\\
\texttt{['|','o','m','l','k']}\\\\
\textbf{MuCoCo Output:}\\
\texttt{['\textasciicircum','o','m','k','e']}\\\\
\end{tabular} &
\xmark & \cmark \\
\midrule

\textbf{Q18} & DeMorgan &
\begin{minipage}{0.5\linewidth}
\vspace{-6pt}
\begin{lstlisting}[language=Python, escapechar=!,]
  def sum_ints_div_by_either_nums(l: List[int]) -> int:
    l = l[$2:$3 + 1]
    return !\textbf{sum([i for i in l if i \% \$0 == 0 or i \% \$1 == 0])}!
\end{lstlisting}
\vspace{-4pt}
\end{minipage} &
\begin{minipage}{0.55\linewidth}
\vspace{-6pt}
\begin{lstlisting}[language=Python, escapechar=!]
  def sum_ints_div_by_either_nums(l: List[int]) -> int:
    l = l[$2:$3 + 1]
    return !\textbf{sum([i for i in l if not (not i \% \$0 == 0 and (not i \% \$1 == 0))])}!
\end{lstlisting}
\vspace{-4pt}
\end{minipage} & 
\begin{tabular}[c]{@{}l@{}}
\textbf{Template Params:}\\
\texttt{  \$0: -4}\\
\texttt{  \$1: 4}\\
\texttt{  \$2: 1}\\
\texttt{  \$3: 6}\\ \\
\textbf{Program Input:}\\ 
\texttt{\lbrack212, -451, -512, 337,}\\
\texttt{486, -442, -160, -422 \rbrack} \\\\
\end{tabular} & 
\begin{tabular}[c]{@{}l@{}}
\textbf{Expected Output:}\\
\texttt{-672} \\\\
\textbf{Turbulence Output:}\\
\texttt{-1114}\\\\
\textbf{MuCoCo Output:}\\
\texttt{-672}\\\\
\end{tabular} &
\begin{tabular}[c]{@{}l@{}}
\textbf{Template Params:}\\
\texttt{  \$0: -6}\\
\texttt{  \$1: 4}\\
\texttt{  \$2: 2}\\
\texttt{  \$3: 4}\\ \\
\textbf{Program Input:}\\ 
\texttt{\lbrack 337, -512, 212,}\\
\texttt{-451, 486, -422\rbrack}\\ \\
\end{tabular} & 
\begin{tabular}[c]{@{}l@{}}
\textbf{Expected Output:}\\
\texttt{698}\\\\
\textbf{Turbulence Output:}\\
\texttt{486}\\\\
\textbf{MuCoCo Output:}\\
\texttt{486}\\\\
\end{tabular} &
\cmark & \xmark \\
\midrule

\textbf{Q3} & Random &
\begin{minipage}{0.56\linewidth}
\vspace{-6pt}
\begin{lstlisting}[language=Python, escapechar=!]
 def !\textbf{all\_pos\_ints\_inclusive}!(l: List[int]) -> List[int]:
   l = l[$0:$1 + 1]
   return [i for i in l if i > 0]
\end{lstlisting}
\vspace{-4pt}
\end{minipage} &
\begin{minipage}{0.55\linewidth}
\vspace{-6pt}
\begin{lstlisting}[language=Python, escapechar=!]
 def !\textbf{eUilVSRGdc}!(l: List[int]) -> List[int]:
   l = l[$0:$1 + 1]
   return [i for i in l if i > 0]
\end{lstlisting}
\vspace{-4pt}
\end{minipage}& 
\begin{tabular}[c]{@{}l@{}}
\textbf{Template Params:}\\
\texttt{  \$0: 0}\\
\texttt{  \$1: 7}\\ \\
\textbf{Program Input:}\\ 
\texttt{\lbrack-451, 486, -531,}\\ 
\texttt{-442, 337, -512,}\\
\texttt{ -160, 212, -422\rbrack}\\ \\
\end{tabular} & 
\begin{tabular}[c]{@{}l@{}}
\textbf{Expected Output:}\\
\texttt{[486, 337, 212]}\\
\textcolor{codegreen}{\texttt{\# List type ``[..]''}}\\\\
\textbf{Turbulence Output:}\\
\texttt{(486, 337, 212)}\\
\textcolor{codegreen}{\texttt{\# Tuple type ``(..)''}}\\\\
\textbf{MuCoCo Output:}\\
\texttt{(486, 337, 212)}\\
\textcolor{codegreen}{\texttt{\# Tuple type ``(..)''}}\\\\
\end{tabular} &
\begin{tabular}[c]{@{}l@{}}
\textbf{Template Params:}\\
\texttt{  \$0: 0}\\
\texttt{  \$1: 1}\\ \\
\textbf{Program Input:}\\ 
\texttt{[-451, -422, 486]}\\ \\
\end{tabular} &
\begin{tabular}[c]{@{}l@{}}
\textbf{Expected Output:}\\
\texttt{[]}\\\\
\textbf{Turbulence Output:}\\
\texttt{[]}\\\\
\textbf{MuCoCo Output:}\\
\texttt{[]}\\\\
\end{tabular} 
& \xmark & \xmark \\
\bottomrule
\end{tabular}
\end{adjustbox}
\end{table*}

\begin{table*}[tb!]
\caption{\centering Impact of varying model confidence thresholds on inconsistency rate (Inc.) and accuracy (Acc.).  Results are reported in the format ``Inc.  (Acc.)'', with accuracy reported in  parenthesis. }
\setlength{\tabcolsep}{5pt}
\renewcommand{\arraystretch}{1.15}
\resizebox{\textwidth}{!}{
\begin{tabular}{c|ccc|ccc|ccc|c}
\toprule
\textbf{Confidence} &
\multicolumn{3}{c|}{\textbf{Gemma}} &
\multicolumn{3}{c|}{\textbf{Qwen}} &
\multicolumn{3}{c|}{\textbf{Llama}} &
\textbf{Agg. Model Inc.} \\ 
\cmidrule(lr){2-4} \cmidrule(lr){5-7} \cmidrule(lr){8-10}
& \textbf{HumanEval} & \textbf{CruxEval} & \textbf{CodeMMLU}
& \textbf{HumanEval} & \textbf{CruxEval} & \textbf{CodeMMLU}
& \textbf{HumanEval} & \textbf{CruxEval} & \textbf{CodeMMLU} & \textbf{Inc.  (Acc.) } \\
\midrule
0.50 &
5.32 (80.44) & 9.42 (73.04) & 31.62 (48.51) &
3.10 (86.89) & 6.38 (81.52) & 26.61 (61.59) &
9.38 (43.37) & 4.65 (91.29) & 33.95 (32.50) &
8.03 (72.79) \\

0.60 &
5.32 (80.44) & 9.42 (73.04) & 31.62 (48.51) &
3.10 (86.89) & 6.38 (81.52) & 26.61 (61.59) &
9.38 (43.37) & 4.65 (91.29) & 33.13 (33.00) &
7.97 (72.83) \\

0.70 &
5.32 (80.44) & 9.21 (73.15) & 31.56 (48.51) &
3.10 (86.89) & 6.38 (81.52) & 25.37 (61.79) &
9.38 (43.37) & 4.65 (91.29) & 22.16 (35.58) &
7.51 (73.21) \\

0.80 & 
2.81 (80.60) & 7.58 (73.70) & 30.21 (48.66) &
0.07 (92.97) & 0.63 (87.47) & 21.07 (64.39) &
0.12 (41.56) & 0.24 (92.97) & 4.70 (38.54) &
3.46 (77.59) \\

0.90 &
1.62 (81.17) & 5.49 (73.75) & 28.53 (49.56) &
0.00 (95.71) & 0.15 (91.98) & 16.67 (66.30) &
0.32 (47.19) & 0.00 (90.29) & 0.00 (59.80) &
3.30 (80.86) \\

0.95 &
1.01 (82.03) & 3.76 (74.74) & 26.91 (49.87) &
0.00 (96.35) & 0.28 (89.37) & 13.35 (71.70) &
0.00 (56.16) & 0.00 (81.88) & 0.00 (72.58) &
3.07 (80.90) \\

0.99 &
0.26 (85.06) & 0.82 (72.60) & 25.28 (47.48) &
0.00 (95.30) & 0.00 (89.94) & 3.73 (78.19) &
0.00 (53.85) & 0.00 (81.63) & 0.00 (93.33) &
1.89 (80.35) \\
\bottomrule
\end{tabular}
}
\label{tab:model-confidence}
\end{table*}

\section{\approach's Scalability to Multiple (Two) Mutations}
\label{sec:multiple-mutation-results}

This experiment examines the scalability of \approach to two simultaneous mutations. 
\autoref{tab:second-order-mutation} highlights our findings.  

Evaluation results show that \textit{second-order mutations reveal inconsistencies at a (\recheck{30.01\%}) higher rate than atomic mutations}.  
\autoref{tab:second-order-mutation} shows that \approach's second-order mutations reveal inconsistency at a \recheck{25.17\%} rate, while atomic mutations have a \recheck{19.36\%} consistency error rate.  
However,  in absolute terms,  \approach's atomic mutations reveal (\recheck{40.38\%}) more consistency errors that its second-order mutations -- \recheck{1269 vs.  904} (\textit{see}  \autoref{tab:second-order-mutation}).  
Overall,  this result demonstrate that \approach scales to multiple mutations and multiple mutations reveal more consistency errors than atomic mutation.  

\begin{result}
\revise{
\approach scales to multiple mutations and they discover consistency errors at a higher rate than atomic mutations.  
}
\end{result}

\section{Impact of Model Confidence on Model Inconsistency}
\label{sec:model-confidence}
This section examines the impact of model confidence on consistency errors found by \approach. 
In particular,  we report how the percentage of consistency errors (aka inconsistency) and model accuracy changes as model confidence thresholds varies between \recheck{0.50 and 0.99}.   
\autoref{tab:model-confidence} reports our findings.  

We observed that \textit{the percentage of consistency errors reduces by up to 76\% as model confidence threshold increases}.  
\autoref{tab:model-confidence} shows that model confidence above 0.50  results in a 8.03\% consistency error rate,  but only 1.89\% consistency error rate is observed at 0.99 confidence threshold.   
Likewise,  we found that model accuracy increases at higher confidence thresholds: 
Increasing model confidence threshold increases model accuracy by up to 7.56\%, on average.  
These results suggest that model confidence improves both consistency and accuracy.   
It also implies that model confidence may be useful in mitigating inconsistency in LLMs, e.g., by only using model outputs with high confidence.  However, we note that at higher confidence threshold,  the number of abstentions (empty LLM  responses) also increases due to the increasing number of LLM outputs that are discarded at high confidence thresholds.  

\begin{result}
   \recheck{
Code inconsistency reduces by up to 76\% as model confidence increases, implying that confidence can serve as a good consistency mitigation method. 
   }
\end{result}

\begin{table*}
\caption{\centering \camcheck{\textsc{Comparison of mutation types supported by \approach against other SOTA frameworks, where \protect\fullcircle\ implies "fully", \protect\halfcircle\ implies "partially" and \protect\emptycircle\ implies "does not" support the specified mutation type.}}}\centering
\resizebox{\textwidth}{!}{
\begin{tabular}{|c|c|c|c|c|c|c|c|c|c|c|c|c|}
\hline
\multirow{2}{*}{Framework}
 & \multicolumn{3}{c|}{Lexical}
 & \multicolumn{2}{c|}{Syntactic}
 & \multicolumn{6}{c|}{Logical}
 & \multirow{2}{*}{\begin{tabular}{c}Natural\\Language\end{tabular}}
\\
\cline{2-12}
 & \begin{tabular}{c}Sequential\\Renaming\end{tabular} & \begin{tabular}{c}Random\\Renaming\end{tabular} & \begin{tabular}{c}Literal\\Format\end{tabular}
 & \begin{tabular}{c}For-to-\\While\end{tabular} & \begin{tabular}{c}For-to-\\Enumerate\end{tabular}
 & \begin{tabular}{c}Boolean\\Literal\end{tabular} & \begin{tabular}{c}De Morgan's\\Law\end{tabular} & \begin{tabular}{c}Commutative\\Reorder\end{tabular} & \begin{tabular}{c}Constant\\Unfold\end{tabular} & \begin{tabular}{c}Constant Unfold\\(Addition)\end{tabular} & \begin{tabular}{c}Constant Unfold\\(Multiplication)\end{tabular}
 &
\\
\hline

\approach (Our approach)
 & \fullcircle & \fullcircle & \fullcircle
 & \fullcircle & \fullcircle
 & \fullcircle & \fullcircle & \fullcircle & \fullcircle & \fullcircle & \fullcircle
 & \emptycircle
\\
\hline

\citep{guan2026benchmarkusefuldynamicbenchmarking}
 & \fullcircle & \fullcircle & \emptycircle
 & \fullcircle & \emptycircle
 & \halfcircle & \emptycircle & \emptycircle & \emptycircle & \fullcircle & \emptycircle
 & \emptycircle
\\
\hline

CodeCrash \citep{NEURIPS2025_aec5e284}
 & \fullcircle & \emptycircle & \emptycircle
 & \emptycircle & \emptycircle
 & \fullcircle & \emptycircle & \emptycircle & \emptycircle & \emptycircle & \emptycircle
 & \fullcircle
\\
\hline

EquiBench \citep{wei-etal-2025-equibench}
 & \halfcircle & \halfcircle & \emptycircle
 & \halfcircle & \halfcircle
 & \emptycircle & \emptycircle & \emptycircle & \emptycircle & \emptycircle & \emptycircle
 & \emptycircle
\\
\hline

CCTest \citep{10172845}
 & \fullcircle & \emptycircle & \emptycircle
 & \emptycircle & \emptycircle
 & \fullcircle & \emptycircle & \emptycircle & \emptycircle & \emptycircle & \emptycircle
 & \emptycircle
\\
\hline

\citep{orvalho2026largelanguagemodelsrobust}
 & \emptycircle & \fullcircle & \emptycircle
 & \fullcircle & \emptycircle
 & \halfcircle & \emptycircle & \emptycircle & \emptycircle & \emptycircle & \emptycircle
 & \emptycircle
\\
\hline

KonTest \citep{rajan-etal-2024-knowledge}
 & \emptycircle & \emptycircle & \emptycircle
 & \emptycircle & \emptycircle
 & \emptycircle & \emptycircle & \emptycircle & \emptycircle & \emptycircle & \emptycircle
 & \fullcircle
\\
\hline

\end{tabular}
}
\label{tab:sota-mucoco-mutation-comparison}
\end{table*}
\section{Effectiveness of Model Ensemble on Mitigating Model Inconsistencies}
\label{sec:model-ensemble}

\camcheck{
This section examines the effectiveness of black-box ensemble strategies for mitigating model inconsistencies.
We adapted the the ensemble strategies in Kontest \citep{rajan-etal-2024-knowledge} for \approach.
The ensemble techniques are as follows:\\
1. \textbf{Input Ensemble}: For each question, the LLM's answers for the mutated questions are first collated before the final answer is decided through majority voting of all mutant answers. 
For questions where a particular mutation does not apply (e.g. for-to-while mutation not applying to a question with no for loop), the mutated question will not count towards the overall votes.\\
2. \textbf{Model Ensemble}: For each question, the answers of each LLM (7 in total) are collated and the final answer is decided through majority voting.\\
3. \textbf{Weighted Model Ensemble}: This strategy is similar to a model ensemble, except that weights are calculated and assigned to each model based on each model's inconsistency rate. 
The equation used for tabulating the weights for each model is shown in \autoref{eq:weighted-ensemble}, where for model $i$, $w_i$ is the model weight, $r_i$ is the model inconsistency rate and $n$ is the number of models (seven in total). 
In this approach, models with lower inconsistency rates are assigned a higher weight, enabling their outputs to contribute more to the overall majority vote, and vice versa for more inconsistent models. 
Additionally, the model weights are normalised to one.\\
4. \textbf{Hybrid Ensemble}: Combining input and model ensembles, we obtain the final answer for each question by majority voting across all models and mutations.
}

\camcheck{It should be noted that when the votes balance out (sum to 0), the ensemble strategy will favour the inconsistent answer for a more modest evaluation.
As a concrete example, let three answers be consistent, and the remaining three answers not. 
The majority voting sums up to zero, which is inconclusive, hence we take the outcome to be inconsistent.}

\camcheck{We would also like to explain the difference in the denominator counts for the various ensemble strategies. 
As an example, let us use HumanEval \citep{chen2021evaluating}, which comprises 164 tasks. 
Assuming all 11 mutations apply to every task and all seven LLMs return valid answers, 
we obtain a total of $164 \times 11 \times 7$ total answers. 
With input ensemble, since the overall answer is aggregated through majority voting of the mutations, we will have $164 \times 7$ total answers. 
With the model ensemble and weighted model ensemble, we aggregate the various LLM answers via majority voting, leaving us with $164 \times 11$ total answers. 
Finally, with the hybrid ensemble, to obtain the aggregated ensemble answer, we apply equal weight to all $7 \times 11$ models and mutations. 
This results in $164$ total answers, hence accounting for the different denominators in the ensemble methods.}

\camcheck{The experiments results are shown in \autoref{tab:ensemble-mitigation}. 
The number of runtime errors are reported separately as they are only accounted for in inconsistencies, not in overall model accuracy, since the model did not actually return an answer within the allotted timeframe. 
The sum of the accuracy denominator and its corresponding total runtime errors will equal the inconsistency denominator.
}
\begin{equation}
w_i = \frac{\dfrac{1}{r_i}}{\sum_{j=1}^{n} \dfrac{1}{r_j}}
\label{eq:weighted-ensemble}
\end{equation}

\begin{result}
   \recheck{
Weighted model ensemble is the most effective ensemble strategy for mitigating model inconsistencies, reducing inconsistency rates by 11.88\%.
   }
\end{result}

\begin{table*}[tb!]
    \caption{\centering 
    \camcheck{Effectiveness of various ensemble mitigation strategies on mitigating model inconsistencies showing Inconsistency Rate (Inc.) and Accuracy (Acc.).}}
    \centering
    \renewcommand{\arraystretch}{1.15}
    \resizebox{\textwidth}{!}{
    \begin{tabular}{c|cc|cc|cc|cc|ccc}
        \textbf{Ensemble} &
        \multicolumn{2}{c|}{\textbf{MCQ}} &
        \multicolumn{2}{c|}{\textbf{Input Pred.}} &
        \multicolumn{2}{c|}{\textbf{Output Pred.}} &
        \multicolumn{2}{c|}{\textbf{Code Gen.}} &
        \multicolumn{3}{c}{\textbf{All Tasks}} \\

        \textbf{Type} & \textbf{Inc.} & \textbf{Acc.} &
          \textbf{Inc.} & \textbf{Acc.} &
          \textbf{Inc.} & \textbf{Acc.} &
          \textbf{Inc.} & \textbf{Acc.} &
          \textbf{Inc.} & \textbf{Acc.} & \textbf{\# Runtime Err.}\\
        \hline

        \textbf{Input Ensemble} &
        20.95 & 65.63 &
        6.22 & 79.68 &
        20.67 & 60.76 &
        30.91 & 51.72 &
        18.02 (6264/34763) & 65.5 (22714/34678) & 85 \\

        \textbf{Model Ensemble} &
        12.54 & 65.35 &
        \textbf{0.29} & 86.42 &
        10.55 & 64.8 &
        15.48 & 67.76 &
        6.85 (1448/21154) & 74.34 (15726/21154) & 0 \\

        \textbf{\begin{tabular}{c}Weighted Model \\ Ensemble\end{tabular}} &
        \textbf{1.69} & 97.75 &
        1.62 & 98.86 &
        \textbf{1.66} & 95.8 &
        \textbf{12.63} & 73.94 &
        \textbf{2.94} (621/21154) & 94.59 (20009/21154) & 0 \\

        \textbf{Hybrid Ensemble} &
        21.47 & 81.6 &
        6.28 & 89.32 &
        15.28 & 66.88 &
        18.13 & 73.78 &
        12.89 (678/5259) & 77.16 (4058/5259) & 0 \\
        \hline

        \textbf{OG \approach} &
        22.35 & 64.04 &
        6.65 & 78.97 &
        18.99 & 62.82 &
        27.15 & 57.66 &
        14.82 (21924/147935) & 69.13 (102092/147680) & 255 \\

    \end{tabular}
    }
\label{tab:ensemble-mitigation}
\end{table*}

\begin{table*}[tb!]
\centering
\caption{\centering \textsc{Inconsistency distance (Inc Dist) across HumanEval, BigCodeBench, and aggregated benchmarks.}}
\renewcommand{\arraystretch}{1.15}
\setlength{\tabcolsep}{6pt}
\resizebox{\textwidth}{!}{%
\begin{tabular}{c|l|ccc|c}
\toprule
 & \textbf{Model} &
\textbf{HumanEval Inc Dist} &
\textbf{BigCodeBench Inc Dist} &
\textbf{Aggregated Inc Dist} &
\textbf{Model Code Gen. Inc} \\
\midrule
\multirow{4}{*}{\rotatebox[origin=c]{90}{\footnotesize\textbf{Open-weight}}}
& Qwen2.5-Coder-14B-Instruct
& 0.301
& 0.192
& 0.204
& 0.410 \\

& Gemma-3-12b-it
& 0.328
& 0.188
& 0.203
& 0.387 \\

& DeepSeek-V3.2-Exp
& 0.043
& 0.146
& 0.135
& 0.194 \\

& LLama-3.1-8B
& \textbf{0.331}
& \textbf{0.223}
& \textbf{0.235}
& \textbf{0.437} \\

\midrule
\multirow{3}{*}{\rotatebox[origin=c]{90}{\footnotesize\textbf{Proprietary}}}
& GPT-5
& 0.015
& 0.103
& 0.094
& 0.126 \\

& GPT-4o
& 0.037
& 0.118
& 0.109
& 0.162 \\

& codestral-2508
& 0.059
& 0.136
& 0.128
& 0.187 \\

\midrule

& Aggregated Open-Source
& 0.251
& 0.187
& 0.194
& 0.357 \\

& Aggregated Proprietary
& 0.037
& 0.119
& 0.110
& 0.158 \\

\bottomrule
\end{tabular}
}
\label{tab:inc-dist-comparison}
\end{table*}

\end{document}